%% file: main.tex
\documentclass[aps, prl, superscriptaddress, twocolumn, 10pt]{revtex4-1}

\include{header}

\begin{document}
	
%%%%%%%%%%%%%%%%%%%%%%%%%%%%%%%%%%%%%%%%%%%%%%%%%%%%%%%%
\title{Micro Black Hole Dark Matter}

\author{Manuel Ettengruber\!\orcidE}
\email{manuel@mpp.mpg.de}
\ThirdAffiliation

\author{Florian K{\"u}hnel\!\orcidC}
\email{fkuehnel@mpp.mpg.de}
\FirstAffiliation
\SecondAffiliation

%%%%%%%%%%%%%%%%%%%%%%%%%%%%%%%%%%%%%
\date{\formatdate{\day }{ \month }{ \year}, \currenttime}

%%%%%%%%%%%%%%%%%%%%%%%%%%%%%%%%%%%%%
\begin{abstract}
The influence of a possible low gravity scale, concretely through the presence of extra dimensions or additional species, on radiation properties of micro black holes is investigated. In particular, the suppression of evaporation through the so-called memory-burden effect is shown to be stronger, occurs earlier and with sharper transition compared to the canonical Planck-scale regime. It is furthermore shown how this affects the possibility of light primordial black hole dark matter, constraints on which may weaken substantially and allow them to be as light as $10^{-5}\.M_{\Prm}$, thereby lying in the particle regime.\\[-5mm]
\end{abstract}

\maketitle

%%%%%%%%%%%%%%%%%%%%%%%%%%%%%%%%%%%%%
\noindent
{\it Introduction\,---}\;An important question in physics is when gravity becomes strong. This question is in fact of fundamental importance for the resolution of the hierarchy problem, whose essence is the inexplicably large separation between the Higgs mass and the Planck mass ($M_{\Prm}$). It has been noted by Dvali {\it et al.}~\cite{Dvali:2007hz, Dvali:2007wp} that the presence of a number of $N$ particle species lowers the strong-gravity scale $M_{\frm}$ as
\vs{-1.5mm}
\begin{subequations}
\begin{equation}
    M_{\frm}
        =
            N^{-1/2}\,
            M_{\Prm}
            \, ,
            \label{eq:masterequation}
\end{equation}
such that sufficiently many additional species would elegantly resolve the hierarchy problem.\footnote{In the Standard Model, the number of particle species is around $100$, and therefore $M_{\Prm}$ and $M_{\frm}$ only differ by an order of magnitude.}

Another, fundamentally related, approach to address the hierarchy problem is by the presence of additional compact spatial dimensions, a prominent example of which is the Arkani-Hamed--Dimopoulos--Dvali (ADD) model~\cite{Arkani-Hamed:1998jmv, Arkani-Hamed:1998sfv, Antoniadis:1998ig}, wherein Eq.~\eqref{eq:masterequation} manifests as
\begin{equation}
    M_{\frm}
        =
            \big(
                2\mspace{1mu}\pi R\.M_{\Prm}
            \big)^{\mspace{-2mu}-n/(n+2)}\,
            M_{\Prm}
            \, ,
            \label{eq:masterequation-Extra-Dimensions}
\end{equation}
\end{subequations}
as pointed out in \cite{Dvali:2007hz}. Here, $n$ is the number of extra dimensions, assumed to be compactified with radius $R$. The latter controls the number of additional (Kaluza--Klein) modes/species present in the effective four-dimensional theory, hence the similarity with the many-species solution of the hierarchy problem mentioned above.

The scale at which gravity becomes strong impacts many facets of black holes, including their evaporation. Since Hawking's seminar paper~\cite{Hawking:1975vcx}, radiation emitted by a black hole has mostly been treated semi-classically, and it was assumed that this regime is only left near $M_{\Prm}$ or possibly $M_{\frm}$. That, however, immediately create a puzzle as first pointed out by Dvali~\cite{Dvali:2015aja}. Imagine that a black hole of mass $M$ and with entropy $S$ evaporates to half of its mass. At this point, it carries a quarter of its original entropy (since the latter is proportional to its area). There is simply no way that the (semiclassically derived) black body emission spectrum could get rid of such an enormous amount of information [cf.~$S \sim 10^{76}\.( M / \Msun )^{2}$]. 

The resolution was proposed by Dvali who suggested that systems of high information capacity, such as black holes, are subject to a universal back-reaction, called ``memory burden" effect~\cite{Dvali:2018xpy}. This not only strongly affects the decay process, but actually stabilises the system~\cite{Dvali:2020wft} and becomes substantial latest by the time the black hole looses half of its mass. The memory-burden effect is a universal property of all systems with high memory capacity compatible with unitarity~\cite{Dvali:2019jjw, Dvali:2019ulr, Dvali:2020wqi} (see Refs.~\cite{Dvali:2021tez, Dvali:2021rlf} for explicit constructions as well as Ref.~\cite{Dvali:2024hsb} for the currently most detailed analysis of entropy-saturated systems).

As first pointed out in Ref.~\cite{Dvali:2020wft}, the memory-burden effect is of particular importance to primordial black holes (PBHs)~\cite{Zeldovich:1967lct, Carr:1974nx, Carr:1975qj} (see Refs.~\cite{Carr:2016drx, Carr:2020xqk, Escriva:2022duf} for reviews). These are black holes which might have formed through non-stellar channels in the early Universe, and would not only be natural dark matter candidates~\cite{Chapline:1975ojl, Carr:2016drx}, but could explain several recent observations, for instance of galaxies beyond redshifts $z = 10$ (cf.~Ref.~\cite{2024MNRAS.533.3222D}), certain microlensing events which cannot have been caused by any of the known astrophysical bodies~\cite{Niikura:2017zjd, Niikura:2019kqi, Hawkins:2022vqo, Hawkins:2023nkq}, or correlations in the source-subtracted cosmic microwave and infrared background which appear to require an early population of non-stellar black holes~\cite{Cappelluti:2012mj, Kashlinsky:2025caq}. These and other hints for the existence of PBHs are summarised in Ref.~\cite{Carr:2023tpt}.

It was put forward in Ref.~\cite{Dvali:2020wft} that memory-burden effect has particularly important implications for PBHs, potentially opening up a new window of light dark matter (below $10^{10}\.\grm$). This was further studied in Refs.~\cite{Alexandre:2024nuo, Dvali:2024hsb, Thoss:2024hsr} as well as in subsequent papers (see, e.g., Refs.~\cite{Boccia:2025hpm, Chianese:2024rsn, Chianese:2025wrk, Calabrese:2025sfh}). Applications concern various astrophysical processes connected to gravitational waves~\cite{Athron:2024fcj, Kohri:2024qpd, Bhaumik:2024qzd, Barman:2024ufm, Barman:2024iht} or the emission of (ultra-)high-energetic particles~\cite{Zantedeschi:2024ram, Chianese:2024rsn, Dondarini:2025ktz, Tan:2025vxp, Chianese:2025wrk, Boccia:2025hpm}. More recently, the smoothness of the effect was taken into account in Refs.~\cite{Montefalcone:2025akm, Dvali:2025ktz}. In earlier studies~\cite{Dvali:2013eja, Dvali:2017eba, Dvali:2020etd, Dvali:2021bsy} based on a microscopic theory of black holes~\cite{Dvali:2011aa}, it has already been understood that species shorten the time of the break-down of the semi-classical regime by a factor $1/N$. A particular application of species to the memory-burden effect was given in Refs.~\cite{Dvali:2021bsy, Alexandre:2024nuo}. While it has already been pointed out in Ref.~\cite{Alexandre:2024nuo} that additional species shorten the transition to the memory-burden phase, the purpose of our work is to elaborate in great detail on this very aspect on the one hand, as well as on that of extra dimensions on the other{\,---\,}both with focus on the possibility that PBHs could be important dark matter components. These topics remain a matter of active research~\cite{Ana}.

This {\it Letter} is structured as follows. We first work out the dependencies of relevant micro black hole quantities on the numbers of additional species and extra dimensions, the influence of which on the memory-burden effect is subsequently studied. We then discuss abundance constraints on micro PBH dark matter, and finally conclude. We use units in which $\hslash = c = k_{\Brm} = 1$.

%%%%%%%%%%%%%%%%%%%%%%%%%%%%%%%%%%%%%
\noindent
{\it Low-Scale Gravity: Species and Extra Dimensions\,---}\;A natural question in the low-scale gravity framework is how black hole physics changes. Of course does the scale $M_{\frm}$ signal where gravity becomes strong and our usual description brakes down but on the other hand for relatively heavy black holes, in particular astrophysical ones, the Einsteinian description should be valid. Therefore one expects another energy scale where the transition from one regime into the other starts taking places. In Refs.~\cite{Dvali:2008fd, Dvali:2008rm} it has been argued that in low-scale gravity theories, one has to distinguish different regimes of black hole physics. Black holes with mass $M$
\begin{equation}
    M_{\frm}
        \ll
            M
        \ll
            \begin{cases}
                \sqrt{N}\.M_{\Prm}\qq
                    &(\textrm{species})\\[1.5mm]
                ( M_{\Prm} R )\.M_{\Prm}\qq
                    &(\textrm{extra dimensions})
            \end{cases}
            \label{eq:Micro-Black-Hole-Regime}
\end{equation}
are in the semiclassical regime where gravity is still classical but their properties are non-Einsteinian.\footnote{The transition length scale for the many-copies model discussed below has been argued to be $R = ( M_{\Prm}^{2} / M_{\frm}^{2} )^{1/n}\.M_{\frm}^{-1}$ in Ref.~\cite{Dvali:2008fd}, exceeding the conservative upper bound in Eq.~\eqref{eq:Micro-Black-Hole-Regime}.} If the mass of a black hole exceeds the upper bound of this interval, it will behave like an ordinary Einsteinian one with the usual Schwarzschild radius (see Refs.~\cite{Dvali:2008rm, Dvali:2008fd}). For the extreme example of $N = 10^{32}$, which is the motivated by the resolution of the hierarchy problem~\cite{Arkani-Hamed:1998jmv, Dvali:2009ne}, the mentioned upper bound becomes $10^{11}\,\grm$ while for the lower bound one has $M_{\frm} \sim 10^{-21}\,\grm$. If it comes close to the scale of $M_{\frm}$, our description of gravity breaks down. 

In the intermediate regime, the Schwarzschild radius can be expressed as~\cite{Argyres:1998qn}
\begin{equation}
    r_{\Srm}^{(\tilde{n})}
        =
            \frac{ a_{\tilde{n}} }
            { M_{\frm} }\mspace{-2mu}
            \left(
                \frac{ M }{ M_{\frm} }
            \right)^{\!1/(1+\tilde{n})}
            ,
            \label{eq:r-S}
\end{equation}
where $\tilde{n}$ and $a_{\tilde{n}}$ are model-dependent parameters. In extra-dimensional theories, the parameter $\tilde{n}$ can be identified with the number $n$ of extra dimensions, and one has~\cite{Argyres:1998qn}
\vs{-1mm}
\begin{equation}
    a_{n}
        =
            \left(
                8\mspace{1mu}\pi^{-(n+1)/2}\,
                \frac{
                \Gamma
                \big[
                    (n+3)/2\mspace{1mu}
                \big]}{n+2}
            \right)^{\!1/(1+n)}
            \, .
            \label{eq:a-n}
\end{equation}
For theories with $N$ species, $\tilde{n}$ depends on the actual realisation of the concrete model~\cite{Dvali:2009ne}, but from phenomenological consideration it is bounded from below as $\tilde{n} > 2$; the parameter $a_{\tilde{n}}$ in Eq.~\eqref{eq:r-S} is unity~\cite{Dvali:2008fd}.

Black hole evaporation is changed accordingly in such theories. The presence of $n$ extra dimensions modify the Hawking temperature as~\cite{Myers:1986un, Myers:1986rx, Argyres:1998qn,Friedlander:2022ttk}
\begin{equation}
    T_{\Hrm}
        =
            \frac{ n + 1 }
            { 4\mspace{1mu}\pi\.r_{\Srm}^{(n)} }
            \, ,
            \label{eq:T-H}
\end{equation}
and the rate of mass-loss becomes~\cite{Friedlander:2022ttk}
\begin{equation}
    \frac{ \d M }{ \d t }\bigg|_{\textrm{extra dim.}}
        =
            -\.\alpha\big( n, T_{\Hrm}^{} \big)\,
            T_{\Hrm}^{2}
            \; .
            \label{eq:dMdt-Extra-Dimensions}            
\end{equation}
Here, $\alpha$ is a numerical coefficient which depends on the thermally-accessible species that can be produced during the evaporation process. In general, $\alpha = \alpha( n,\.T_{\Hrm} )$, but when $T_{\Hrm}$ is much larger than the masses $m_{j}$ of the species, $\alpha$ saturates to a constant value. Depending on the number of extra dimensions and $T_{\Hrm}$, $\alpha( n,\.T_{\Hrm} )$ typically ranges between unity and $100$, and can be calculated via~\cite{Friedlander:2022ttk}
\begin{equation}
    \frac{ \d M }{ \d t }\bigg|_{\textrm{extra dim.}}
        =
            \frac{ 1 }{ 2\mspace{1mu}\pi }
            \sum_{j} \xi^{}_{j}
            \left(
                \frac{ 1 }{ r_{\Srm}^{(n)} }
            \right)^{\!2}
        \equiv
            -\mspace{1mu}\alpha(n,\.T_{\Hrm})\;
            T_{\Hrm}^{2}
            \; ,
            \label{eq:alpha-def}
\end{equation}
where the parameters $\xi_{i}$, incorporate the integrated particle-emission spectrum and are accounting for the degrees of freedom of the particles. Their functional shape can be approximated by~\cite{Friedlander:2022ttk} 
\begin{equation}
    \xi_{i}
        =
            \xi_{i,\mspace{1mu}0}\.
            \erm^{
                -\mspace{1mu}b_{j}\mspace{1.5mu}
                ( m_{j} / T_{\Hrm} )^{c_{j}}
            }
            \, ,
            \label{eq:xi-i}
\end{equation}
with $b_{j} = 0.3$ and $c_{j} = 1.3$ for Standard Model particles, and $\xi_{j,\mspace{1mu}0}$ is the value obtained using $m_{j} = 0$ (see Ref.~\cite{Friedlander:2022ttk} for details). Note that $\alpha$ receives additive contributions:
\begin{equation}
    \alpha
        =
            \alpha_{\textrm{scalars}}
            + \alpha_{\textrm{gauge bosons}}
            + \alpha_{\textrm{fermions}}
            + \alpha_{\textrm{gravitons}}
            \, .
            \label{eq:alpha-sum}
\end{equation}
Restricting to Standard Model degrees of freedom, one obtains for the graviton emission in the high-temperature regime (wherein $T_{\Hrm} > m_{j}$), $\alpha_{\textrm{gravitons}} = 0.1,\,\ldots,\,14.4\%$ for $n = 0,\,\ldots,\,6$, respectively, as shown in Refs.~\cite{Cardoso:2005mh, Friedlander:2022ttk}. 

Analogously, black hole evaporation is also affected by the presence of additional species~\cite{Dvali:2007hz, Dvali:2007wp, Alexandre:2024nuo} into which the hole would radiate democratically in the Einsteinian regime. For semi-classical black holes and the exemplary case of $N$ only-gravitationally-coupled copies of the Standard Model, the fraction of the energy which gets emitted into our copy as compared to all of the others is~\cite{Dvali:2008fd}
\begin{equation}
    \frac{ E_{{\rm BH}\,\longrightarrow\,\textrm{our sector}} }
    { E_{{\rm BH}\,\longrightarrow\,\textrm{all sectors}} }
        \sim
            \mspace{-2mu}
            \left(
                \frac{ M_{\frm} }{ M }
            \right)^{\!\tilde{n}/(1+\tilde{n})}\
            \, ,
            \label{eq:E-Ratio}
\end{equation}
and the evaporation rate gets modified to 
\begin{equation}
    \frac{ \d M }{ \d t }\bigg|_{\textrm{species}}
        =
            -\mspace{1.5mu}M_{\frm}^{2}\mspace{-2mu}
            \left(
                \frac{ M_{\frm} }{ M }
            \right)^{\!( 2 - \tilde{n} ) /( 1 + \tilde{n})}
            \label{eq:dMdt-Species}
            \, .
\end{equation}

For the memory-burden effect it is of particular interest of how the entropy changes within these two frameworks. For the specific ADD case, the $(4+n)$-dimensional entropy, $S_{n}$, of a Schwarzschildian black hole can be expressed as~\cite{Argyres:1998qn}
\begin{align}
    S_{\textrm{extra dim.}}
        =
            S^{(4)}\mspace{-1mu}
            \left(
                \frac{ R }{ r_{\Srm} }
            \right)^{\!n/(n+1)}\mspace{-9mu}
        \sim
            S^{(4)}\mspace{-1mu}
            \left(
                \frac{ M_{\Prm}^{2n+2} }
                { M_{\frm}^{n+2} M_{}^{n} }
            \right)^{\!1/(n+1)}\mspace{-12mu}
            ,
            \label{eq:S-Extra-Dimensions}
            \\[-5mm]
            \notag
\end{align}
where $S^{(4)} \sim M_{\Prm}^{2}\,r_{\Srm}^{2}$ is the usual four-dimensional entropy, and $r_{\Srm} \sim M / M_{\Prm}^{2}$ is the canonical Schwarzschild radius. With $M_{\frm} / M_{\Prm} = 10^{-15}$ and $n = 2$, which is a typical choice for the ADD model, one gets $S_{n} \simeq 10^{14}\.S^{(4)}$ for a black hole of mass $10^{4}\,\grm$.

The effect of additional species manifests in a similar way, wherefore one obtains
\begin{equation}
     S_{\textrm{species}}
        \sim S^{(4)}
            \left(
                \frac{ M_{\Prm} }{ M }
            \right)^{\!2\mspace{1mu}\tilde{n}/(\tilde{n}+1)}\!           
                N
            ^{1/(\tilde{n}+1)}
            \, .
            \label{eq:S-Species}
\end{equation}
Concretely, for a black hole of mass $M_{\Prm}$ with $\tilde{n} = 3$, its entropy is eight orders of magnitude larger than in the four-dimensional case.

In the convoluted case of $N$ additional extra-dimen-sional species, as considered in Ref.~\cite{Dvali:2010vm}, Eq.~\eqref{eq:masterequation} becomes
\begin{equation}
    M_{\frm}
        \rightarrow
            N^{-1/(2+n)}\,M_{\frm}
            \, , 
\end{equation}
where $M_{\frm}$ on the right-hand side is calculated using Eq.~\eqref{eq:masterequation-Extra-Dimensions}. Notice that the formula for the Schwarzschild radius in such theories, Eq.~\eqref{eq:r-S}, remains the same (up to factors of order one) and one can calculate the resulting entropy of such black holes with the final expression given in Eq.~\eqref{eq:S-Extra-Dimensions}.

Under the assumption that black hole evaporation is essentially entirely semiclassical ('SC'), the lifetime $\tau|_{\textrm{SC}} \sim M_{\Prm}^{-1}\.( M / M_{\Prm} )^{3}$ of a four-dimensional Schwarz-schildian black hole, becomes in ADD scenarios~\cite{Argyres:1998qn}
\begin{equation}
    \tau_{\textrm{extra dim.}}\Big|_{\rm SC}
        = 
            \frac{ C_{\rm ADD} }
            { M_{\frm} }\mspace{-2mu}
            \left(
                \frac{ M }{ M_{\frm} }
            \right)^{\!(n+3)/(n+1)}
            ,
            \label{eq:tau-Extra-Dimensions}
\end{equation}
where $C_{\rm ADD}$ can be expressed as
\begin{equation}
    C_{\rm ADD}
        =
            \frac{ 16\mspace{1mu}\pi^{2}\.a_{n}^{2} }
            { \alpha\mspace{1.5mu}(n+1)(n+3) }
            \, .
            \label{eq:C-ADD}
\end{equation}
On the other hand, for $N$ additional species, the semiclassical black hole lifetime reads~\cite{Dvali:2008fd}
\begin{equation}
    \tau_{\textrm{species}}\Big|_{\rm SC}
        =
            \frac{ C^{(N)} }{ M_{\frm} }\mspace{-2mu}
            \left(
                \frac{ M }{ M_{\frm} }
            \right)^{\!3/(\tilde{n}+1)}
            ,
            \label{eq:tau-Species}
\end{equation}
with $C^{(N)}$ being given by
\begin{equation}
    C^{(N)}
        =
           \frac{ \tilde{n} + 1 }{ 3 }
            \, .
            \label{eq:C-N}
\end{equation}
As we will discuss below, there are strong arguments against the assumption of the validity of the semiclassical regime throughout most of the evaporation process, due to sizeable{\,---\,}lifetime prolonging{\,---\,}quantum effects.
\vs{-3mm}

%%%%%%%%%%%%%%%%%%%%%%%%%%%%%%%%%%%%%
\noindent
{\it Memory Burden\,---}\;As mentioned, a practical consequence of the memory-burden effect is its induced suppression of the Hawking evaporation rate, leading to light black hole (quasi) relict states which could be cosmologically long-lived and would thereby constitute natural dark matter candidates (see, e.g., Refs.~\cite{Dvali:2020wft, Alexandre:2024nuo, Dvali:2024hsb, Thoss:2024hsr}). The original memory-burden ('MB') formula for the suppression of the black hole evaporation rate is (cf.~Ref.~\cite{Dvali:2020wft})
\begin{align}
    \frac{ \d M }{ \d t }\bigg|_{\rm MB}
        =
            \frac{ 1 }{ S^{k} }\.
            \frac{ \d M }{ \d t }
            \bigg|_{\rm SC}
            \, ,
            \label{eq:dMdt-Memory-Burden}
\end{align}
which applies after the evaporation of a fraction $1 - q$ of the initial black hole mass. The exponent $k$ remains to be determined, but it has been argued to be integer~\cite{Dvali:2024hsb}.

Equation~\eqref{eq:dMdt-Memory-Burden} makes it apparent how low-scale gravity theories influence the evaporation behaviour in the memory-burden phase due to the change in entropy [cf.~Eqs.~(\ref{eq:S-Extra-Dimensions}, \ref{eq:S-Species})]. In extra-dimensional frameworks, the entropy for a micro black hole with mass in the range~\eqref{eq:Micro-Black-Hole-Regime} is strictly larger than that of an ordinary Einsteinian one. This means that as soon as a black hole exits the semi-classical phase, the memory-burden effect will be significantly enhanced and the evaporation slows down.

In the many-species case, this behaviour is not so strict. As soon the semi-classical regime has been entered, the ratio $M_{\Prm} / M$ that appears in Eq.~\eqref{eq:S-Species} can be small enough to push the prefactor that describes the deviation from the ordinary entropy to a smaller value. In other words, the radius shrinks in the semi-classical regime so much that the entropy, which scales with the area of the hole, is decreasing accordingly. As soon the black hole looses more mass due to radiation and shrinks even further, it will enter a phase where the enhancement due to the presence of a large number of species dominates and $S_{\textrm{species}} > S^{(4)}$, like in the extra-dimensional case. Therefore, both cases can severely impact how the burden of memory affects black hole evaporation in such theories.

Until recently, in phenomenological studies, the transition from the semiclassical phase has been approximated with a sharp transition. Recent work~\cite{Montefalcone:2025akm, Dvali:2025ktz} has investigated the impact of a smooth transition with width $\delta$ on PBH abundance constraints. For specific black hole prototype models, Ref.~\cite{Dvali:2025ktz} has provided analytical expressions for $q$ and $\delta$, which read
\vs{1mm}
\begin{subequations}
\begin{align}
    q
        &\simeq
            \big(
                p^{2}\mspace{1mu}
                S
            \big)^{-1/[2(p-1)]}
            \, ,
            \label{eq:q-S-p}
            \displaybreak[1]
            \\[2.5mm]
    \delta
        &=
             \frac{ q }{ (p-1)\mspace{1mu}\ln S }
        \simeq
            \frac{ 
            \big(
                p^{2}\mspace{1mu}
                S
            \big)^{-1/[2(p-1)]} }{ (p-1)\mspace{1mu}\ln S }
            \qq
            (p \ne 1)
            \, ,
            \label{eq:delta-S-p}
\end{align}
\end{subequations}
respectively, with the integer $p \geq 1$ being a model parameter.\footnote{The transition occurs essentially instantly for $p = 1$, but also for $p = 2$, one has $q(p = 2) \simeq 1 / \sqrt{S}$, which is extremely early, given that $S \simeq 10^{30}\,( M / 10^{10}\,\grm )^{2}$.} Using Eqs.~(\ref{eq:q-S-p},{\color{midblue}b}), we can express $q$ and $\delta$ for the cases of extra dimensions and additional species, using Eq.~\eqref{eq:S-Extra-Dimensions} and Eq.~\eqref{eq:S-Species}, respectively. This yields the scaling behaviours (ignoring subdominant logarithms)
\begin{subequations}
\begin{align}
    q_{\textrm{extra dim.}}
        &\propto
            \big(
                M_{\Prm}^{2}\,r_{\Srm}^{2}
            \big)^{\mspace{-1.5mu}-1/[2(p-1)]}\,
            \bigg(
                \frac{ r_{\Srm} }{ R }
            \bigg)^{\!n/[2(n+1)(p-1)]}
%\xrightarrow[n\.\rightarrow\.\infty]{}
%            \left(
%                \frac{ r_{\Srm} }{ R }
%            \right)^{\!1/2(p-1)}
            ,
            \\[3mm]
    q_{\textrm{species}}
        &\propto
            \left(
                \frac{ M_{\Prm} }{ M }
            \right)^{\!1/[(\tilde{n}+1)(p-1)]}
            N^{-1/[2(\tilde{n}+1)(p-1)]}
%\xrightarrow[N\.\rightarrow\.\infty]{}
%           0
            \, .
\end{align}
\end{subequations}
The corresponding relations for $\delta$ can be expressed in terms of $q$ via Eq.~\eqref{eq:delta-S-p}.
%Concretely, for the choices $n = \FK{...}$ and $N = \FK{...}$, we find $q_{n}^{(N)} / q = \FK{...}$, $\delta_{n}^{(N)} / \delta = \FK{...}$, while in the presence of many additional species, $q_{n}^{(N \ra \infty)} / q = \delta_{n}^{(N \ra \infty)} / \delta = 0$, leading to an instantaneous transition. 
Below, we show the associated phenomenological implications for micro black holes as potential dark matter candidates.

%%%%%%%%%%%%%%%%%%%%%%%%%%%%%%%%%%%%%
\noindent
{\it Micro Primordial Black Holes Dark Matter\,---}\;Let us now turn to the implications of our conceptual considerations above on the possibility that such micro black holes could be dark matter.

In the ADD case, already without employing the memory-burden effect, one can see from Eq.~\eqref{eq:tau-Extra-Dimensions} that the lifetime of micro black holes in a higher-dimensional scenario is usually larger than that of an ordinary four-dimensional one, simply because $M_{\Prm}$ gets replaced by $M_{\frm} \ll M_{\Prm}$ in the denominator of Eq.~\eqref{eq:tau-Extra-Dimensions}, thereby enlarging $\tau_{\textrm{extra dim.}}$. For example, the semiclassical lifetime of a four-dimensional black hole of mass $10^{10}\,\grm$ is around $400\,\srm$, while for $n = 2$ and $M_{\frm} = 10\,\TeV$ it becomes already $10^{21}\,\srm$, far exceeding the age of the Universe $t_{0} \approx 10^{17}\,\srm$. This fact triggered research of higher-dimensional black holes as dark matter candidates. For two popular models with compactified extra-dimensions like we are investigating here, see Refs.~\cite{Conley:2006jg, Friedlander:2022ttk}; for the ADD case and the connected so-called Dark Dimension scenarios, this has been discussed in Refs.~\cite{Anchordoqui:2022txe, Anchordoqui:2022txe, Anchordoqui:2022tgp, Anchordoqui:2024akj, Anchordoqui:2024dxu, Anchordoqui:2024jkn, Anchordoqui:2024tdj, Anchordoqui:2025nmb, Anchordoqui:2025opy,
Anchordoqui:2025xug}.

Of course, the memory-burden effect will further increase the lifetime of such black holes. The importance of higher dimensions in this context was first discussed in Ref.~\cite{Anchordoqui:2024dxu}. From recent work~\cite{Dvali:2025ktz}, it has been become clear that the transition to the memory-burden phase appears much earlier than Page's time as opposed to what was originally anticipated. As pointed out in Eq.~\eqref{eq:q-S-p}, the time when the effect sets is decreased by inverse powers of the entropy. As it is clear from Eqs.~\eqref{eq:S-Extra-Dimensions} and~\eqref{eq:S-Species}, this is of fundamental importance for theories with low-scale gravity which yield a significantly enlarged entropy of such black holes.

So what is the smallest black hole an ADD scenario with the memory-burden effect would allow which has not yet evaporated? From the Eqs.~\eqref{eq:r-S} and~\eqref{eq:S-Extra-Dimensions} it becomes clear that the strongest effect is expected for $n = 2$. Then, by taking $k = 2$, we get for the lifetime:
\begin{equation}
    \tau_{\textrm{ADD}}^{n=2}
        \sim
            a_{n}\,
            \frac{ M^{4} }{ M_{\frm}^{5} }
            \left(
                \frac{ M }{ M_{\frm} }
            \right)^{\!1/3}
            .
\end{equation}
In turn, by requiring that such holes should live at least as long as the age of the Universe, yields 
\begin{equation}
    M
        \gtrsim
            10^{14}\,\GeV
        \approx
            10^{-10}\,\grm
            \, ,
\end{equation}
being five orders of magnitude below the Planck mass, so {\it the masses of these objects overlap with that of heavy particles}.

When it comes to the bounds on the PBH abundance stemming from Big Bang nucleosynthesis (BBN) and cosmic microwave background (CMB) radiation, the situation is different. For instance, in Ref.~\cite{Friedlander:2022ttk} it was shown that (semiclassically) the bounds do not vanish in an ADD scenario even though they get weakened. Hence, the memory-burden effect is particularly impactful here for three reasons: Firstly, the increase of entropy~\eqref{eq:S-Extra-Dimensions} suppresses the evaporation rate~\eqref{eq:dMdt-Memory-Burden}. Secondly, the transition from the semiclassical to the memory-burden phase appears dramatically earlier [lower $q$-values, cf.~Eq.~\eqref{eq:q-S-p}], and, thirdly, its width narrows (smaller $\delta$-values) as can be inferred from Eq.~\eqref{eq:delta-S-p}. A detailed quantitative study on the modification of the PBH abundance constraints will be presented in a subsequent publication~\cite{Ettengruber-Constraints-2025}.

In the case of many species, i.e.~$N \gg 1$, the situation inverted in a sense: Instead of having an extremely low bound on the abundance of black hole which have already evaporation by now, while still having the constraints stemming from BBN or the CMB, here, the latter is not affected by black hole evaporation. As apparent from Eq.~\eqref{eq:E-Ratio}, the holes will mostly evaporate into dark species, leaving therefore only a small fraction of Standard Model evaporation products which would interfere with BBN or the CMB. Consequently, evaporation bounds on the abundance of such black holes vanish, leaving only non-evaporation bounds (currently constraining mass ranges above $M \leq 10^{23}\,\grm$).

On the other, the bound on the abundance of black hole which have already evaporated by now gets strengthened by the number of available species. For $\tilde{n} = 3$ and $k = 2$ one has
\begin{equation}
    \tau_{\textrm{species}}^{\tilde{n}=3}
        \sim
            \left(
                \frac{ M }{ M_{\frm} }
            \right)^{\!9/4}
            \frac{ 1 }{ \sqrt{M\.M_{\frm} }}
            \, ,
\end{equation}
which leads to a lower bound of 
\begin{equation}
    M
        \gtrsim
            10^{30}\,\GeV
        \approx
            10^{5}\,\grm
            \, .
\end{equation}
Interestingly, this is just on the edge of the light PBH dark matter mass window which was originally investigated in the literature (cf.~Refs.~\cite{Alexandre:2024nuo, Thoss:2024hsr}). For theories with extra species, the memory-burden effect is essential to make the black holes sufficiently longe-lived to be dark matter candidates.
\vs{2mm}

%%%%%%%%%%%%%%%%%%%%%%%%%%%%%%%%%%%%%
\noindent
{\it Conclusion\,---}\;In this {\it Letter} we have investigated the influence of extra dimensions and additional species on the evaporation dynamics of micro black holes with respect to the memory-burden effect. We have shown that for large parameter ranges, their lifetime is substantially prolonged, in turn opening and enlarging the dark matter window which was closed by semi-classical Hawking evaporation bounds. In a subsequent phenomenological study~\cite{Ettengruber-Constraints-2025}, we will quantify the abundance constraints on primordial black holes.
\vs{2mm}

%%%%%%%%%%%%%%%%%%%%%%%%%%%%%%%%%%%%%
\noindent
{\it Acknowledgments\,---}\;It is a pleasure to thank Gia Dvali for critical reading of the manuscript and for invaluable discussions. Stimulating conversations with Gabriele Montefalcone are also acknowledged. We also thank Luis Anchordoqui, Ignatios Antoniadis, Fabio Iocco, Dieter L{\"u}st and Michael Zantedeschi for referential suggestions. The work of M.E.~was supported by ANR grant ANR-23-CE31-0024 EUHiggs.

%%%%%%%%%%%%%%%%%%%%%%%%%%%%%%%%%%%%%
\bibliography{refs}

\end{document}

%% file: header.tex
\usepackage{amsmath}	% AMS Math Package
\usepackage{amsthm}		% Theorem Formatting
\usepackage{amssymb}	% Math symbols such as \mathbb
\usepackage{array}
\usepackage{datetime}
\usepackage{graphicx}
\usepackage{color}
\usepackage{verbatim}
\usepackage{bm}
\usepackage{braket}
\usepackage{natbib}
\usepackage{cancel}
\usepackage{xcolor}
\usepackage{slashed}
\usepackage{braket}
\usepackage{amsfonts} 
\smallskipamount
\usepackage{feynmp}
\usepackage{letltxmacro} % closed roots
\usepackage[colorlinks=true, citecolor=midblue, linkcolor=midblue, urlcolor=midblue]{hyperref}

\usepackage{setspace}
\usepackage[normalem]{ulem}

\usepackage{tikz,xcolor}
\definecolor{lime}{HTML}{A6CE39}
\DeclareRobustCommand{\orcidicon}{
	\begin{tikzpicture}
	\draw[lime, fill=lime] (0,0) 
	circle [radius=0.16] 
	node[white] {{\fontfamily{qag}\selectfont \tiny ID}};
	\draw[white, fill=white] (-0.0625,0.095) 
	circle [radius=0.007];
	\end{tikzpicture}
	\hspace{-2mm}
}
\foreach \x in {A, ..., Z}{
	\expandafter\xdef\csname orcid\x\endcsname{\noexpand\href{https://orcid.org/\csname orcidauthor\x\endcsname}{\noexpand\orcidicon}}
}

\settimeformat{ampmtime}

\definecolor{grey}{rgb}{0.4,0.4,0.4}
\definecolor{dullmagenta}{rgb}{0.4,0,0.4}
\definecolor{darkblue}{rgb}{0,0,0.4}
\definecolor{midblue}{rgb}{0,0,0.5}
\definecolor{midred}{rgb}{0.5,0,0}
\definecolor{orange}{rgb}{1,0.5,0}
\definecolor{lightbrown}{rgb}{0.75,0.5,0.25}
\definecolor{tan}{cmyk}{0.14,0.42,0.56,0}
\definecolor{djunglegreen}{cmyk}{0.99,0,0.52,0}
\definecolor{lightgreen}{rgb}{0,1,0}
\definecolor{olivegreen}{cmyk}{0.64,0,0.95,0.40}
\definecolor{midgreen}{rgb}{0.0,0.675,0.0}
\definecolor{darkgreen}{rgb}{0,0.5,0}

\newcommand{\qq}{\qquad}

\newcommand{\vs}{\vspace}

\renewcommand{\.}{\hspace{0.5mm}}

\newcommand{\Brm}{\ensuremath{\mathrm{B}}}

\newcommand{\Hrm}{\ensuremath{\mathrm{H}}}

\newcommand{\Prm}{\ensuremath{\mathrm{P}}}

\newcommand{\Srm}{\ensuremath{\mathrm{S}}}

\newcommand{\erm}{\ensuremath{\mathrm{e}}}
\newcommand{\frm}{\ensuremath{\mathrm{f}}}
\newcommand{\grm}{\ensuremath{\mathrm{g}}}

\newcommand{\srm}{\ensuremath{\mathrm{s}}}

\renewcommand{\d}{\ensuremath{\mathrm{d}}}
\newcommand{\GeV}{\ensuremath{\mathrm{GeV}}}
\newcommand{\TeV}{\ensuremath{\mathrm{TeV}}}

\def\Msun{M_\odot}

\settimeformat{ampmtime}

\makeatletter
\let\oldr@@t\r@@t
\def\r@@t#1#2{%
\setbox0=\hbox{$\oldr@@t#1{#2\,}$}\dimen0=\ht0
\advance\dimen0-0.2\ht0
\setbox2=\hbox{\vrule height\ht0 depth -\dimen0}%
{\box0\lower0.4pt\box2}}
\LetLtxMacro{\oldsqrt}{\sqrt}
\renewcommand*{\sqrt}[2][\ ]{\oldsqrt[#1]{#2}}
\makeatother

\newcommand{\FirstAffiliation}{\affiliation{
	Arnold Sommerfeld Center,
	Ludwig-Maximilians-Universit{\"a}t,
	Theresienstra{\ss}e 37,
	80333 M{\"u}nchen,
	Germany}}
 
\newcommand{\SecondAffiliation}{\affiliation{
	Max-Planck-Institut f{\"u}r Physik,
	Boltzmannstr.~8, 
	85748 Garching,
	Germany}}

\newcommand{\ThirdAffiliation}{\affiliation{Universit{\'e} Paris--Saclay, CNRS, CEA, Institut de Physique Theorique, 91191, Gif-sur-Yvette, France
}}

%% file: main.bbl
%merlin.mbs apsrev4-1.bst 2010-07-25 4.21a (PWD, AO, DPC) hacked
%Control: key (0)
%Control: author (8) initials jnrlst
%Control: editor formatted (1) identically to author
%Control: production of article title (-1) disabled
%Control: page (0) single
%Control: year (1) truncated
%Control: production of eprint (0) enabled
\begin{thebibliography}{75}%
\makeatletter
\providecommand \@ifxundefined [1]{%
 \@ifx{#1\undefined}
}%
\providecommand \@ifnum [1]{%
 \ifnum #1\expandafter \@firstoftwo
 \else \expandafter \@secondoftwo
 \fi
}%
\providecommand \@ifx [1]{%
 \ifx #1\expandafter \@firstoftwo
 \else \expandafter \@secondoftwo
 \fi
}%
\providecommand \natexlab [1]{#1}%
\providecommand \enquote  [1]{``#1''}%
\providecommand \bibnamefont  [1]{#1}%
\providecommand \bibfnamefont [1]{#1}%
\providecommand \citenamefont [1]{#1}%
\providecommand \href@noop [0]{\@secondoftwo}%
\providecommand \href [0]{\begingroup \@sanitize@url \@href}%
\providecommand \@href[1]{\@@startlink{#1}\@@href}%
\providecommand \@@href[1]{\endgroup#1\@@endlink}%
\providecommand \@sanitize@url [0]{\catcode `\\12\catcode `\$12\catcode `\&12\catcode `\#12\catcode `\^12\catcode `\_12\catcode `\%12\relax}%
\providecommand \@@startlink[1]{}%
\providecommand \@@endlink[0]{}%
\providecommand \url  [0]{\begingroup\@sanitize@url \@url }%
\providecommand \@url [1]{\endgroup\@href {#1}{\urlprefix }}%
\providecommand \urlprefix  [0]{URL }%
\providecommand \Eprint [0]{\href }%
\providecommand \doibase [0]{http://dx.doi.org/}%
\providecommand \selectlanguage [0]{\@gobble}%
\providecommand \bibinfo  [0]{\@secondoftwo}%
\providecommand \bibfield  [0]{\@secondoftwo}%
\providecommand \translation [1]{[#1]}%
\providecommand \BibitemOpen [0]{}%
\providecommand \bibitemStop [0]{}%
\providecommand \bibitemNoStop [0]{.\EOS\space}%
\providecommand \EOS [0]{\spacefactor3000\relax}%
\providecommand \BibitemShut  [1]{\csname bibitem#1\endcsname}%
\let\auto@bib@innerbib\@empty
%</preamble>
\bibitem [{\citenamefont {Dvali}(2010{\natexlab{a}})}]{Dvali:2007hz}%
  \BibitemOpen
  \bibfield  {author} {\bibinfo {author} {\bibfnamefont {G.}~\bibnamefont {Dvali}},\ }\href {\doibase 10.1002/prop.201000009} {\bibfield  {journal} {\bibinfo  {journal} {Fortsch. Phys.}\ }\textbf {\bibinfo {volume} {58}},\ \bibinfo {pages} {528} (\bibinfo {year} {2010}{\natexlab{a}})},\ \Eprint {http://arxiv.org/abs/0706.2050} {arXiv:0706.2050 [hep-th]} \BibitemShut {NoStop}%
\bibitem [{\citenamefont {Dvali}\ and\ \citenamefont {Redi}(2008)}]{Dvali:2007wp}%
  \BibitemOpen
  \bibfield  {author} {\bibinfo {author} {\bibfnamefont {G.}~\bibnamefont {Dvali}}\ and\ \bibinfo {author} {\bibfnamefont {M.}~\bibnamefont {Redi}},\ }\href {\doibase 10.1103/PhysRevD.77.045027} {\bibfield  {journal} {\bibinfo  {journal} {Phys. Rev. D}\ }\textbf {\bibinfo {volume} {77}},\ \bibinfo {pages} {045027} (\bibinfo {year} {2008})},\ \Eprint {http://arxiv.org/abs/0710.4344} {arXiv:0710.4344 [hep-th]} \BibitemShut {NoStop}%
\bibitem [{Note1()}]{Note1}%
  \BibitemOpen
  \bibinfo {note} {In the Standard Model, the number of particle species is around $100$, and therefore $M_{\protect \ensuremath {\protect \mathrm {P}}}$ and $M_{\protect \ensuremath {\protect \mathrm {f}}}$ only differ by an order of magnitude.}\BibitemShut {Stop}%
\bibitem [{\citenamefont {Arkani-Hamed}\ \emph {et~al.}(1998)\citenamefont {Arkani-Hamed}, \citenamefont {Dimopoulos},\ and\ \citenamefont {Dvali}}]{Arkani-Hamed:1998jmv}%
  \BibitemOpen
  \bibfield  {author} {\bibinfo {author} {\bibfnamefont {N.}~\bibnamefont {Arkani-Hamed}}, \bibinfo {author} {\bibfnamefont {S.}~\bibnamefont {Dimopoulos}}, \ and\ \bibinfo {author} {\bibfnamefont {G.~R.}\ \bibnamefont {Dvali}},\ }\href {\doibase 10.1016/S0370-2693(98)00466-3} {\bibfield  {journal} {\bibinfo  {journal} {Phys. Lett. B}\ }\textbf {\bibinfo {volume} {429}},\ \bibinfo {pages} {263} (\bibinfo {year} {1998})},\ \Eprint {http://arxiv.org/abs/hep-ph/9803315} {arXiv:hep-ph/9803315} \BibitemShut {NoStop}%
\bibitem [{\citenamefont {Arkani-Hamed}\ \emph {et~al.}(1999)\citenamefont {Arkani-Hamed}, \citenamefont {Dimopoulos},\ and\ \citenamefont {Dvali}}]{Arkani-Hamed:1998sfv}%
  \BibitemOpen
  \bibfield  {author} {\bibinfo {author} {\bibfnamefont {N.}~\bibnamefont {Arkani-Hamed}}, \bibinfo {author} {\bibfnamefont {S.}~\bibnamefont {Dimopoulos}}, \ and\ \bibinfo {author} {\bibfnamefont {G.~R.}\ \bibnamefont {Dvali}},\ }\href {\doibase 10.1103/PhysRevD.59.086004} {\bibfield  {journal} {\bibinfo  {journal} {Phys. Rev. D}\ }\textbf {\bibinfo {volume} {59}},\ \bibinfo {pages} {086004} (\bibinfo {year} {1999})},\ \Eprint {http://arxiv.org/abs/hep-ph/9807344} {arXiv:hep-ph/9807344} \BibitemShut {NoStop}%
\bibitem [{\citenamefont {Antoniadis}\ \emph {et~al.}(1998)\citenamefont {Antoniadis}, \citenamefont {Arkani-Hamed}, \citenamefont {Dimopoulos},\ and\ \citenamefont {Dvali}}]{Antoniadis:1998ig}%
  \BibitemOpen
  \bibfield  {author} {\bibinfo {author} {\bibfnamefont {I.}~\bibnamefont {Antoniadis}}, \bibinfo {author} {\bibfnamefont {N.}~\bibnamefont {Arkani-Hamed}}, \bibinfo {author} {\bibfnamefont {S.}~\bibnamefont {Dimopoulos}}, \ and\ \bibinfo {author} {\bibfnamefont {G.~R.}\ \bibnamefont {Dvali}},\ }\href {\doibase 10.1016/S0370-2693(98)00860-0} {\bibfield  {journal} {\bibinfo  {journal} {Phys. Lett. B}\ }\textbf {\bibinfo {volume} {436}},\ \bibinfo {pages} {257} (\bibinfo {year} {1998})},\ \Eprint {http://arxiv.org/abs/hep-ph/9804398} {arXiv:hep-ph/9804398} \BibitemShut {NoStop}%
\bibitem [{\citenamefont {Hawking}(1975)}]{Hawking:1975vcx}%
  \BibitemOpen
  \bibfield  {author} {\bibinfo {author} {\bibfnamefont {S.~W.}\ \bibnamefont {Hawking}},\ }\href {\doibase 10.1007/BF02345020} {\bibfield  {journal} {\bibinfo  {journal} {Commun. Math. Phys.}\ }\textbf {\bibinfo {volume} {43}},\ \bibinfo {pages} {199} (\bibinfo {year} {1975})},\ \bibinfo {note} {[Erratum: Commun.Math.Phys. 46, 206 (1976)]}\BibitemShut {NoStop}%
\bibitem [{\citenamefont {Dvali}(2016)}]{Dvali:2015aja}%
  \BibitemOpen
  \bibfield  {author} {\bibinfo {author} {\bibfnamefont {G.}~\bibnamefont {Dvali}},\ }\href {\doibase 10.1002/prop.201500096} {\bibfield  {journal} {\bibinfo  {journal} {Fortsch. Phys.}\ }\textbf {\bibinfo {volume} {64}},\ \bibinfo {pages} {106} (\bibinfo {year} {2016})},\ \Eprint {http://arxiv.org/abs/1509.04645} {arXiv:1509.04645 [hep-th]} \BibitemShut {NoStop}%
\bibitem [{\citenamefont {Dvali}(2018)}]{Dvali:2018xpy}%
  \BibitemOpen
  \bibfield  {author} {\bibinfo {author} {\bibfnamefont {G.}~\bibnamefont {Dvali}},\ }\href@noop {} {\  (\bibinfo {year} {2018})},\ \Eprint {http://arxiv.org/abs/1810.02336} {arXiv:1810.02336 [hep-th]} \BibitemShut {NoStop}%
\bibitem [{\citenamefont {Dvali}\ \emph {et~al.}(2020)\citenamefont {Dvali}, \citenamefont {Eisemann}, \citenamefont {Michel},\ and\ \citenamefont {Zell}}]{Dvali:2020wft}%
  \BibitemOpen
  \bibfield  {author} {\bibinfo {author} {\bibfnamefont {G.}~\bibnamefont {Dvali}}, \bibinfo {author} {\bibfnamefont {L.}~\bibnamefont {Eisemann}}, \bibinfo {author} {\bibfnamefont {M.}~\bibnamefont {Michel}}, \ and\ \bibinfo {author} {\bibfnamefont {S.}~\bibnamefont {Zell}},\ }\href {\doibase 10.1103/PhysRevD.102.103523} {\bibfield  {journal} {\bibinfo  {journal} {Phys. Rev. D}\ }\textbf {\bibinfo {volume} {102}},\ \bibinfo {pages} {103523} (\bibinfo {year} {2020})},\ \Eprint {http://arxiv.org/abs/2006.00011} {arXiv:2006.00011 [hep-th]} \BibitemShut {NoStop}%
\bibitem [{\citenamefont {Dvali}(2021{\natexlab{a}})}]{Dvali:2019jjw}%
  \BibitemOpen
  \bibfield  {author} {\bibinfo {author} {\bibfnamefont {G.}~\bibnamefont {Dvali}},\ }\href {\doibase 10.1002/prop.202000090} {\bibfield  {journal} {\bibinfo  {journal} {Fortsch. Phys.}\ }\textbf {\bibinfo {volume} {69}},\ \bibinfo {pages} {2000090} (\bibinfo {year} {2021}{\natexlab{a}})},\ \Eprint {http://arxiv.org/abs/1906.03530} {arXiv:1906.03530 [hep-th]} \BibitemShut {NoStop}%
\bibitem [{\citenamefont {Dvali}(2021{\natexlab{b}})}]{Dvali:2019ulr}%
  \BibitemOpen
  \bibfield  {author} {\bibinfo {author} {\bibfnamefont {G.}~\bibnamefont {Dvali}},\ }\href {\doibase 10.1002/prop.202000091} {\bibfield  {journal} {\bibinfo  {journal} {Fortsch. Phys.}\ }\textbf {\bibinfo {volume} {69}},\ \bibinfo {pages} {2000091} (\bibinfo {year} {2021}{\natexlab{b}})},\ \Eprint {http://arxiv.org/abs/1907.07332} {arXiv:1907.07332 [hep-th]} \BibitemShut {NoStop}%
\bibitem [{\citenamefont {Dvali}(2021{\natexlab{c}})}]{Dvali:2020wqi}%
  \BibitemOpen
  \bibfield  {author} {\bibinfo {author} {\bibfnamefont {G.}~\bibnamefont {Dvali}},\ }\href {\doibase 10.1007/JHEP03(2021)126} {\bibfield  {journal} {\bibinfo  {journal} {JHEP}\ }\textbf {\bibinfo {volume} {03}},\ \bibinfo {pages} {126} (\bibinfo {year} {2021}{\natexlab{c}})},\ \Eprint {http://arxiv.org/abs/2003.05546} {arXiv:2003.05546 [hep-th]} \BibitemShut {NoStop}%
\bibitem [{\citenamefont {Dvali}\ \emph {et~al.}(2022)\citenamefont {Dvali}, \citenamefont {Kaikov},\ and\ \citenamefont {Berm\'udez}}]{Dvali:2021tez}%
  \BibitemOpen
  \bibfield  {author} {\bibinfo {author} {\bibfnamefont {G.}~\bibnamefont {Dvali}}, \bibinfo {author} {\bibfnamefont {O.}~\bibnamefont {Kaikov}}, \ and\ \bibinfo {author} {\bibfnamefont {J.~S.~V.}\ \bibnamefont {Berm\'udez}},\ }\href {\doibase 10.1103/PhysRevD.105.056013} {\bibfield  {journal} {\bibinfo  {journal} {Phys. Rev. D}\ }\textbf {\bibinfo {volume} {105}},\ \bibinfo {pages} {056013} (\bibinfo {year} {2022})},\ \Eprint {http://arxiv.org/abs/2112.00551} {arXiv:2112.00551 [hep-th]} \BibitemShut {NoStop}%
\bibitem [{\citenamefont {Dvali}\ and\ \citenamefont {Sakhelashvili}(2022)}]{Dvali:2021rlf}%
  \BibitemOpen
  \bibfield  {author} {\bibinfo {author} {\bibfnamefont {G.}~\bibnamefont {Dvali}}\ and\ \bibinfo {author} {\bibfnamefont {O.}~\bibnamefont {Sakhelashvili}},\ }\href {\doibase 10.1103/PhysRevD.105.065014} {\bibfield  {journal} {\bibinfo  {journal} {Phys. Rev. D}\ }\textbf {\bibinfo {volume} {105}},\ \bibinfo {pages} {065014} (\bibinfo {year} {2022})},\ \Eprint {http://arxiv.org/abs/2111.03620} {arXiv:2111.03620 [hep-th]} \BibitemShut {NoStop}%
\bibitem [{\citenamefont {Dvali}\ \emph {et~al.}(2024)\citenamefont {Dvali}, \citenamefont {Valbuena-Berm\'udez},\ and\ \citenamefont {Zantedeschi}}]{Dvali:2024hsb}%
  \BibitemOpen
  \bibfield  {author} {\bibinfo {author} {\bibfnamefont {G.}~\bibnamefont {Dvali}}, \bibinfo {author} {\bibfnamefont {J.~S.}\ \bibnamefont {Valbuena-Berm\'udez}}, \ and\ \bibinfo {author} {\bibfnamefont {M.}~\bibnamefont {Zantedeschi}},\ }\href {\doibase 10.1103/PhysRevD.110.056029} {\bibfield  {journal} {\bibinfo  {journal} {Phys. Rev. D}\ }\textbf {\bibinfo {volume} {110}},\ \bibinfo {pages} {056029} (\bibinfo {year} {2024})},\ \Eprint {http://arxiv.org/abs/2405.13117} {arXiv:2405.13117 [hep-th]} \BibitemShut {NoStop}%
\bibitem [{\citenamefont {Zel'dovich}\ and\ \citenamefont {Novikov}(1967)}]{Zeldovich:1967lct}%
  \BibitemOpen
  \bibfield  {author} {\bibinfo {author} {\bibfnamefont {Y.~B.}\ \bibnamefont {Zel'dovich}}\ and\ \bibinfo {author} {\bibfnamefont {I.~D.}\ \bibnamefont {Novikov}},\ }\href@noop {} {\bibfield  {journal} {\bibinfo  {journal} {Sov. Astron.}\ }\textbf {\bibinfo {volume} {10}},\ \bibinfo {pages} {602} (\bibinfo {year} {1967})}\BibitemShut {NoStop}%
\bibitem [{\citenamefont {Carr}\ and\ \citenamefont {Hawking}(1974)}]{Carr:1974nx}%
  \BibitemOpen
  \bibfield  {author} {\bibinfo {author} {\bibfnamefont {B.~J.}\ \bibnamefont {Carr}}\ and\ \bibinfo {author} {\bibfnamefont {S.~W.}\ \bibnamefont {Hawking}},\ }\href {\doibase 10.1093/mnras/168.2.399} {\bibfield  {journal} {\bibinfo  {journal} {Mon. Not. Roy. Astron. Soc.}\ }\textbf {\bibinfo {volume} {168}},\ \bibinfo {pages} {399} (\bibinfo {year} {1974})}\BibitemShut {NoStop}%
\bibitem [{\citenamefont {Carr}(1975)}]{Carr:1975qj}%
  \BibitemOpen
  \bibfield  {author} {\bibinfo {author} {\bibfnamefont {B.~J.}\ \bibnamefont {Carr}},\ }\href {\doibase 10.1086/153853} {\bibfield  {journal} {\bibinfo  {journal} {Astrophys. J.}\ }\textbf {\bibinfo {volume} {201}},\ \bibinfo {pages} {1} (\bibinfo {year} {1975})}\BibitemShut {NoStop}%
\bibitem [{\citenamefont {Carr}\ \emph {et~al.}(2016)\citenamefont {Carr}, \citenamefont {Kuhnel},\ and\ \citenamefont {Sandstad}}]{Carr:2016drx}%
  \BibitemOpen
  \bibfield  {author} {\bibinfo {author} {\bibfnamefont {B.}~\bibnamefont {Carr}}, \bibinfo {author} {\bibfnamefont {F.}~\bibnamefont {Kuhnel}}, \ and\ \bibinfo {author} {\bibfnamefont {M.}~\bibnamefont {Sandstad}},\ }\href {\doibase 10.1103/PhysRevD.94.083504} {\bibfield  {journal} {\bibinfo  {journal} {Phys. Rev. D}\ }\textbf {\bibinfo {volume} {94}},\ \bibinfo {pages} {083504} (\bibinfo {year} {2016})},\ \Eprint {http://arxiv.org/abs/1607.06077} {arXiv:1607.06077 [astro-ph.CO]} \BibitemShut {NoStop}%
\bibitem [{\citenamefont {Carr}\ and\ \citenamefont {Kuhnel}(2020)}]{Carr:2020xqk}%
  \BibitemOpen
  \bibfield  {author} {\bibinfo {author} {\bibfnamefont {B.}~\bibnamefont {Carr}}\ and\ \bibinfo {author} {\bibfnamefont {F.}~\bibnamefont {Kuhnel}},\ }\href {\doibase 10.1146/annurev-nucl-050520-125911} {\bibfield  {journal} {\bibinfo  {journal} {Ann. Rev. Nucl. Part. Sci.}\ }\textbf {\bibinfo {volume} {70}},\ \bibinfo {pages} {355} (\bibinfo {year} {2020})},\ \Eprint {http://arxiv.org/abs/2006.02838} {arXiv:2006.02838 [astro-ph.CO]} \BibitemShut {NoStop}%
\bibitem [{\citenamefont {Escriv\`a}\ \emph {et~al.}(2022)\citenamefont {Escriv\`a}, \citenamefont {Kuhnel},\ and\ \citenamefont {Tada}}]{Escriva:2022duf}%
  \BibitemOpen
  \bibfield  {author} {\bibinfo {author} {\bibfnamefont {A.}~\bibnamefont {Escriv\`a}}, \bibinfo {author} {\bibfnamefont {F.}~\bibnamefont {Kuhnel}}, \ and\ \bibinfo {author} {\bibfnamefont {Y.}~\bibnamefont {Tada}},\ }\href {\doibase 10.1016/B978-0-32-395636-9.00012-8} {\  (\bibinfo {year} {2022}),\ 10.1016/B978-0-32-395636-9.00012-8},\ \Eprint {http://arxiv.org/abs/2211.05767} {arXiv:2211.05767 [astro-ph.CO]} \BibitemShut {NoStop}%
\bibitem [{\citenamefont {Chapline}(1975)}]{Chapline:1975ojl}%
  \BibitemOpen
  \bibfield  {author} {\bibinfo {author} {\bibfnamefont {G.~F.}\ \bibnamefont {Chapline}},\ }\href {\doibase 10.1038/253251a0} {\bibfield  {journal} {\bibinfo  {journal} {Nature}\ }\textbf {\bibinfo {volume} {253}},\ \bibinfo {pages} {251} (\bibinfo {year} {1975})}\BibitemShut {NoStop}%
\bibitem [{\citenamefont {{Donnan}}\ \emph {et~al.}(2024)\citenamefont {{Donnan}}, \citenamefont {{McLure}}, \citenamefont {{Dunlop}}, \citenamefont {{McLeod}}, \citenamefont {{Magee}}, \citenamefont {{Arellano-C{\'o}rdova}}, \citenamefont {{Barrufet}}, \citenamefont {{Begley}}, \citenamefont {{Bowler}}, \citenamefont {{Carnall}}, \citenamefont {{Cullen}}, \citenamefont {{Ellis}}, \citenamefont {{Fontana}}, \citenamefont {{Illingworth}}, \citenamefont {{Grogin}}, \citenamefont {{Hamadouche}}, \citenamefont {{Koekemoer}}, \citenamefont {{Liu}}, \citenamefont {{Mason}}, \citenamefont {{Santini}},\ and\ \citenamefont {{Stanton}}}]{2024MNRAS.533.3222D}%
  \BibitemOpen
  \bibfield  {author} {\bibinfo {author} {\bibfnamefont {C.~T.}\ \bibnamefont {{Donnan}}}, \bibinfo {author} {\bibfnamefont {R.~J.}\ \bibnamefont {{McLure}}}, \bibinfo {author} {\bibfnamefont {J.~S.}\ \bibnamefont {{Dunlop}}}, \bibinfo {author} {\bibfnamefont {D.~J.}\ \bibnamefont {{McLeod}}}, \bibinfo {author} {\bibfnamefont {D.}~\bibnamefont {{Magee}}}, \bibinfo {author} {\bibfnamefont {K.~Z.}\ \bibnamefont {{Arellano-C{\'o}rdova}}}, \bibinfo {author} {\bibfnamefont {L.}~\bibnamefont {{Barrufet}}}, \bibinfo {author} {\bibfnamefont {R.}~\bibnamefont {{Begley}}}, \bibinfo {author} {\bibfnamefont {R.~A.~A.}\ \bibnamefont {{Bowler}}}, \bibinfo {author} {\bibfnamefont {A.~C.}\ \bibnamefont {{Carnall}}}, \bibinfo {author} {\bibfnamefont {F.}~\bibnamefont {{Cullen}}}, \bibinfo {author} {\bibfnamefont {R.~S.}\ \bibnamefont {{Ellis}}}, \bibinfo {author} {\bibfnamefont {A.}~\bibnamefont {{Fontana}}}, \bibinfo {author} {\bibfnamefont {G.~D.}\ \bibnamefont {{Illingworth}}}, \bibinfo {author} {\bibfnamefont {N.~A.}\
  \bibnamefont {{Grogin}}}, \bibinfo {author} {\bibfnamefont {M.~L.}\ \bibnamefont {{Hamadouche}}}, \bibinfo {author} {\bibfnamefont {A.~M.}\ \bibnamefont {{Koekemoer}}}, \bibinfo {author} {\bibfnamefont {F.~Y.}\ \bibnamefont {{Liu}}}, \bibinfo {author} {\bibfnamefont {C.}~\bibnamefont {{Mason}}}, \bibinfo {author} {\bibfnamefont {P.}~\bibnamefont {{Santini}}}, \ and\ \bibinfo {author} {\bibfnamefont {T.~M.}\ \bibnamefont {{Stanton}}},\ }\href {\doibase 10.1093/mnras/stae2037} {\bibfield  {journal} {\bibinfo  {journal} {Mon. Not. R. Astron. Soc}\ }\textbf {\bibinfo {volume} {533}},\ \bibinfo {pages} {3222} (\bibinfo {year} {2024})},\ \Eprint {http://arxiv.org/abs/2403.03171} {arXiv:2403.03171 [astro-ph.GA]} \BibitemShut {NoStop}%
\bibitem [{\citenamefont {Niikura}\ \emph {et~al.}(2019{\natexlab{a}})\citenamefont {Niikura} \emph {et~al.}}]{Niikura:2017zjd}%
  \BibitemOpen
  \bibfield  {author} {\bibinfo {author} {\bibfnamefont {H.}~\bibnamefont {Niikura}} \emph {et~al.},\ }\href {\doibase 10.1038/s41550-019-0723-1} {\bibfield  {journal} {\bibinfo  {journal} {Nature Astron.}\ }\textbf {\bibinfo {volume} {3}},\ \bibinfo {pages} {524} (\bibinfo {year} {2019}{\natexlab{a}})},\ \Eprint {http://arxiv.org/abs/1701.02151} {arXiv:1701.02151 [astro-ph.CO]} \BibitemShut {NoStop}%
\bibitem [{\citenamefont {Niikura}\ \emph {et~al.}(2019{\natexlab{b}})\citenamefont {Niikura}, \citenamefont {Takada}, \citenamefont {Yokoyama}, \citenamefont {Sumi},\ and\ \citenamefont {Masaki}}]{Niikura:2019kqi}%
  \BibitemOpen
  \bibfield  {author} {\bibinfo {author} {\bibfnamefont {H.}~\bibnamefont {Niikura}}, \bibinfo {author} {\bibfnamefont {M.}~\bibnamefont {Takada}}, \bibinfo {author} {\bibfnamefont {S.}~\bibnamefont {Yokoyama}}, \bibinfo {author} {\bibfnamefont {T.}~\bibnamefont {Sumi}}, \ and\ \bibinfo {author} {\bibfnamefont {S.}~\bibnamefont {Masaki}},\ }\href {\doibase 10.1103/PhysRevD.99.083503} {\bibfield  {journal} {\bibinfo  {journal} {Phys. Rev. D}\ }\textbf {\bibinfo {volume} {99}},\ \bibinfo {pages} {083503} (\bibinfo {year} {2019}{\natexlab{b}})},\ \Eprint {http://arxiv.org/abs/1901.07120} {arXiv:1901.07120 [astro-ph.CO]} \BibitemShut {NoStop}%
\bibitem [{\citenamefont {Hawkins}(2022)}]{Hawkins:2022vqo}%
  \BibitemOpen
  \bibfield  {author} {\bibinfo {author} {\bibfnamefont {M.~R.~S.}\ \bibnamefont {Hawkins}},\ }\href {\doibase 10.1093/mnras/stac863} {\bibfield  {journal} {\bibinfo  {journal} {Mon. Not. Roy. Astron. Soc.}\ }\textbf {\bibinfo {volume} {512}},\ \bibinfo {pages} {5706} (\bibinfo {year} {2022})},\ \Eprint {http://arxiv.org/abs/2204.09143} {arXiv:2204.09143 [astro-ph.CO]} \BibitemShut {NoStop}%
\bibitem [{\citenamefont {Hawkins}(2023)}]{Hawkins:2023nkq}%
  \BibitemOpen
  \bibfield  {author} {\bibinfo {author} {\bibfnamefont {M.~R.~S.}\ \bibnamefont {Hawkins}},\ }\href {\doibase 10.1093/mnras/stad3346} {\bibfield  {journal} {\bibinfo  {journal} {Mon. Not. Roy. Astron. Soc.}\ }\textbf {\bibinfo {volume} {527}},\ \bibinfo {pages} {2393} (\bibinfo {year} {2023})},\ \Eprint {http://arxiv.org/abs/2311.08915} {arXiv:2311.08915 [astro-ph.CO]} \BibitemShut {NoStop}%
\bibitem [{\citenamefont {Cappelluti}\ \emph {et~al.}(2013)\citenamefont {Cappelluti}, \citenamefont {Kashlinsky}, \citenamefont {Arendt}, \citenamefont {Comastri}, \citenamefont {Fazio}, \citenamefont {Finoguenov}, \citenamefont {Hasinger}, \citenamefont {Mather}, \citenamefont {Miyaji},\ and\ \citenamefont {Moseley}}]{Cappelluti:2012mj}%
  \BibitemOpen
  \bibfield  {author} {\bibinfo {author} {\bibfnamefont {N.}~\bibnamefont {Cappelluti}}, \bibinfo {author} {\bibfnamefont {A.}~\bibnamefont {Kashlinsky}}, \bibinfo {author} {\bibfnamefont {R.~G.}\ \bibnamefont {Arendt}}, \bibinfo {author} {\bibfnamefont {A.}~\bibnamefont {Comastri}}, \bibinfo {author} {\bibfnamefont {G.~G.}\ \bibnamefont {Fazio}}, \bibinfo {author} {\bibfnamefont {A.}~\bibnamefont {Finoguenov}}, \bibinfo {author} {\bibfnamefont {G.}~\bibnamefont {Hasinger}}, \bibinfo {author} {\bibfnamefont {J.~C.}\ \bibnamefont {Mather}}, \bibinfo {author} {\bibfnamefont {T.}~\bibnamefont {Miyaji}}, \ and\ \bibinfo {author} {\bibfnamefont {S.~H.}\ \bibnamefont {Moseley}},\ }\href {\doibase 10.1088/0004-637X/769/1/68} {\bibfield  {journal} {\bibinfo  {journal} {Astrophys. J.}\ }\textbf {\bibinfo {volume} {769}},\ \bibinfo {pages} {68} (\bibinfo {year} {2013})},\ \Eprint {http://arxiv.org/abs/1210.5302} {arXiv:1210.5302 [astro-ph.CO]} \BibitemShut {NoStop}%
\bibitem [{\citenamefont {Kashlinsky}\ \emph {et~al.}(2025)\citenamefont {Kashlinsky}, \citenamefont {Arendt}, \citenamefont {Ashby}, \citenamefont {Kruk},\ and\ \citenamefont {Odegard}}]{Kashlinsky:2025caq}%
  \BibitemOpen
  \bibfield  {author} {\bibinfo {author} {\bibfnamefont {A.}~\bibnamefont {Kashlinsky}}, \bibinfo {author} {\bibfnamefont {R.~G.}\ \bibnamefont {Arendt}}, \bibinfo {author} {\bibfnamefont {M.~L.~N.}\ \bibnamefont {Ashby}}, \bibinfo {author} {\bibfnamefont {J.}~\bibnamefont {Kruk}}, \ and\ \bibinfo {author} {\bibfnamefont {N.}~\bibnamefont {Odegard}},\ }\href {\doibase 10.3847/2041-8213/adad5e} {\bibfield  {journal} {\bibinfo  {journal} {Astrophys. J. Lett.}\ }\textbf {\bibinfo {volume} {980}},\ \bibinfo {pages} {L12} (\bibinfo {year} {2025})},\ \Eprint {http://arxiv.org/abs/2501.17751} {arXiv:2501.17751 [astro-ph.CO]} \BibitemShut {NoStop}%
\bibitem [{\citenamefont {Carr}\ \emph {et~al.}(2024)\citenamefont {Carr}, \citenamefont {Clesse}, \citenamefont {Garcia-Bellido}, \citenamefont {Hawkins},\ and\ \citenamefont {Kuhnel}}]{Carr:2023tpt}%
  \BibitemOpen
  \bibfield  {author} {\bibinfo {author} {\bibfnamefont {B.}~\bibnamefont {Carr}}, \bibinfo {author} {\bibfnamefont {S.}~\bibnamefont {Clesse}}, \bibinfo {author} {\bibfnamefont {J.}~\bibnamefont {Garcia-Bellido}}, \bibinfo {author} {\bibfnamefont {M.}~\bibnamefont {Hawkins}}, \ and\ \bibinfo {author} {\bibfnamefont {F.}~\bibnamefont {Kuhnel}},\ }\href {\doibase 10.1016/j.physrep.2023.11.005} {\bibfield  {journal} {\bibinfo  {journal} {Phys. Rept.}\ }\textbf {\bibinfo {volume} {1054}},\ \bibinfo {pages} {1} (\bibinfo {year} {2024})},\ \Eprint {http://arxiv.org/abs/2306.03903} {arXiv:2306.03903 [astro-ph.CO]} \BibitemShut {NoStop}%
\bibitem [{\citenamefont {Alexandre}\ \emph {et~al.}(2024)\citenamefont {Alexandre}, \citenamefont {Dvali},\ and\ \citenamefont {Koutsangelas}}]{Alexandre:2024nuo}%
  \BibitemOpen
  \bibfield  {author} {\bibinfo {author} {\bibfnamefont {A.}~\bibnamefont {Alexandre}}, \bibinfo {author} {\bibfnamefont {G.}~\bibnamefont {Dvali}}, \ and\ \bibinfo {author} {\bibfnamefont {E.}~\bibnamefont {Koutsangelas}},\ }\href {\doibase 10.1103/PhysRevD.110.036004} {\bibfield  {journal} {\bibinfo  {journal} {Phys. Rev. D}\ }\textbf {\bibinfo {volume} {110}},\ \bibinfo {pages} {036004} (\bibinfo {year} {2024})},\ \Eprint {http://arxiv.org/abs/2402.14069} {arXiv:2402.14069 [hep-ph]} \BibitemShut {NoStop}%
\bibitem [{\citenamefont {Thoss}\ \emph {et~al.}(2024)\citenamefont {Thoss}, \citenamefont {Burkert},\ and\ \citenamefont {Kohri}}]{Thoss:2024hsr}%
  \BibitemOpen
  \bibfield  {author} {\bibinfo {author} {\bibfnamefont {V.}~\bibnamefont {Thoss}}, \bibinfo {author} {\bibfnamefont {A.}~\bibnamefont {Burkert}}, \ and\ \bibinfo {author} {\bibfnamefont {K.}~\bibnamefont {Kohri}},\ }\href {\doibase 10.1093/mnras/stae1098} {\bibfield  {journal} {\bibinfo  {journal} {Mon. Not. Roy. Astron. Soc.}\ }\textbf {\bibinfo {volume} {532}},\ \bibinfo {pages} {451} (\bibinfo {year} {2024})},\ \Eprint {http://arxiv.org/abs/2402.17823} {arXiv:2402.17823 [astro-ph.CO]} \BibitemShut {NoStop}%
\bibitem [{\citenamefont {Boccia}\ and\ \citenamefont {Iocco}(2025)}]{Boccia:2025hpm}%
  \BibitemOpen
  \bibfield  {author} {\bibinfo {author} {\bibfnamefont {A.}~\bibnamefont {Boccia}}\ and\ \bibinfo {author} {\bibfnamefont {F.}~\bibnamefont {Iocco}},\ }\href@noop {} {\  (\bibinfo {year} {2025})},\ \Eprint {http://arxiv.org/abs/2502.19245} {arXiv:2502.19245 [astro-ph.HE]} \BibitemShut {NoStop}%
\bibitem [{\citenamefont {Chianese}\ \emph {et~al.}(2025)\citenamefont {Chianese}, \citenamefont {Boccia}, \citenamefont {Iocco}, \citenamefont {Miele},\ and\ \citenamefont {Saviano}}]{Chianese:2024rsn}%
  \BibitemOpen
  \bibfield  {author} {\bibinfo {author} {\bibfnamefont {M.}~\bibnamefont {Chianese}}, \bibinfo {author} {\bibfnamefont {A.}~\bibnamefont {Boccia}}, \bibinfo {author} {\bibfnamefont {F.}~\bibnamefont {Iocco}}, \bibinfo {author} {\bibfnamefont {G.}~\bibnamefont {Miele}}, \ and\ \bibinfo {author} {\bibfnamefont {N.}~\bibnamefont {Saviano}},\ }\href {\doibase 10.1103/PhysRevD.111.063036} {\bibfield  {journal} {\bibinfo  {journal} {Phys. Rev. D}\ }\textbf {\bibinfo {volume} {111}},\ \bibinfo {pages} {063036} (\bibinfo {year} {2025})},\ \Eprint {http://arxiv.org/abs/2410.07604} {arXiv:2410.07604 [astro-ph.HE]} \BibitemShut {NoStop}%
\bibitem [{\citenamefont {Chianese}(2025)}]{Chianese:2025wrk}%
  \BibitemOpen
  \bibfield  {author} {\bibinfo {author} {\bibfnamefont {M.}~\bibnamefont {Chianese}},\ }\href@noop {} {\  (\bibinfo {year} {2025})},\ \Eprint {http://arxiv.org/abs/2504.03838} {arXiv:2504.03838 [astro-ph.HE]} \BibitemShut {NoStop}%
\bibitem [{\citenamefont {Calabrese}\ \emph {et~al.}(2025)\citenamefont {Calabrese}, \citenamefont {Chianese},\ and\ \citenamefont {Saviano}}]{Calabrese:2025sfh}%
  \BibitemOpen
  \bibfield  {author} {\bibinfo {author} {\bibfnamefont {R.}~\bibnamefont {Calabrese}}, \bibinfo {author} {\bibfnamefont {M.}~\bibnamefont {Chianese}}, \ and\ \bibinfo {author} {\bibfnamefont {N.}~\bibnamefont {Saviano}},\ }\href {\doibase 10.1103/PhysRevD.111.083008} {\bibfield  {journal} {\bibinfo  {journal} {Phys. Rev. D}\ }\textbf {\bibinfo {volume} {111}},\ \bibinfo {pages} {083008} (\bibinfo {year} {2025})},\ \Eprint {http://arxiv.org/abs/2501.06298} {arXiv:2501.06298 [hep-ph]} \BibitemShut {NoStop}%
\bibitem [{\citenamefont {Athron}\ \emph {et~al.}(2025)\citenamefont {Athron}, \citenamefont {Chianese}, \citenamefont {Datta}, \citenamefont {Samanta},\ and\ \citenamefont {Saviano}}]{Athron:2024fcj}%
  \BibitemOpen
  \bibfield  {author} {\bibinfo {author} {\bibfnamefont {P.}~\bibnamefont {Athron}}, \bibinfo {author} {\bibfnamefont {M.}~\bibnamefont {Chianese}}, \bibinfo {author} {\bibfnamefont {S.}~\bibnamefont {Datta}}, \bibinfo {author} {\bibfnamefont {R.}~\bibnamefont {Samanta}}, \ and\ \bibinfo {author} {\bibfnamefont {N.}~\bibnamefont {Saviano}},\ }\href {\doibase 10.1088/1475-7516/2025/05/005} {\bibfield  {journal} {\bibinfo  {journal} {JCAP}\ }\textbf {\bibinfo {volume} {05}},\ \bibinfo {pages} {005} (\bibinfo {year} {2025})},\ \Eprint {http://arxiv.org/abs/2411.19286} {arXiv:2411.19286 [astro-ph.CO]} \BibitemShut {NoStop}%
\bibitem [{\citenamefont {Kohri}\ \emph {et~al.}(2025)\citenamefont {Kohri}, \citenamefont {Terada},\ and\ \citenamefont {Yanagida}}]{Kohri:2024qpd}%
  \BibitemOpen
  \bibfield  {author} {\bibinfo {author} {\bibfnamefont {K.}~\bibnamefont {Kohri}}, \bibinfo {author} {\bibfnamefont {T.}~\bibnamefont {Terada}}, \ and\ \bibinfo {author} {\bibfnamefont {T.~T.}\ \bibnamefont {Yanagida}},\ }\href {\doibase 10.1103/PhysRevD.111.063543} {\bibfield  {journal} {\bibinfo  {journal} {Phys. Rev. D}\ }\textbf {\bibinfo {volume} {111}},\ \bibinfo {pages} {063543} (\bibinfo {year} {2025})},\ \Eprint {http://arxiv.org/abs/2409.06365} {arXiv:2409.06365 [astro-ph.CO]} \BibitemShut {NoStop}%
\bibitem [{\citenamefont {Bhaumik}\ \emph {et~al.}(2024)\citenamefont {Bhaumik}, \citenamefont {Haque}, \citenamefont {Jain},\ and\ \citenamefont {Lewicki}}]{Bhaumik:2024qzd}%
  \BibitemOpen
  \bibfield  {author} {\bibinfo {author} {\bibfnamefont {N.}~\bibnamefont {Bhaumik}}, \bibinfo {author} {\bibfnamefont {M.~R.}\ \bibnamefont {Haque}}, \bibinfo {author} {\bibfnamefont {R.~K.}\ \bibnamefont {Jain}}, \ and\ \bibinfo {author} {\bibfnamefont {M.}~\bibnamefont {Lewicki}},\ }\href {\doibase 10.1007/JHEP10(2024)142} {\bibfield  {journal} {\bibinfo  {journal} {JHEP}\ }\textbf {\bibinfo {volume} {10}},\ \bibinfo {pages} {142} (\bibinfo {year} {2024})},\ \Eprint {http://arxiv.org/abs/2409.04436} {arXiv:2409.04436 [astro-ph.CO]} \BibitemShut {NoStop}%
\bibitem [{\citenamefont {Barman}\ \emph {et~al.}(2024{\natexlab{a}})\citenamefont {Barman}, \citenamefont {Loho},\ and\ \citenamefont {Zapata}}]{Barman:2024ufm}%
  \BibitemOpen
  \bibfield  {author} {\bibinfo {author} {\bibfnamefont {B.}~\bibnamefont {Barman}}, \bibinfo {author} {\bibfnamefont {K.}~\bibnamefont {Loho}}, \ and\ \bibinfo {author} {\bibfnamefont {O.}~\bibnamefont {Zapata}},\ }\href {\doibase 10.1088/1475-7516/2024/10/065} {\bibfield  {journal} {\bibinfo  {journal} {JCAP}\ }\textbf {\bibinfo {volume} {10}},\ \bibinfo {pages} {065} (\bibinfo {year} {2024}{\natexlab{a}})},\ \Eprint {http://arxiv.org/abs/2409.05953} {arXiv:2409.05953 [gr-qc]} \BibitemShut {NoStop}%
\bibitem [{\citenamefont {Barman}\ \emph {et~al.}(2024{\natexlab{b}})\citenamefont {Barman}, \citenamefont {Haque},\ and\ \citenamefont {Zapata}}]{Barman:2024iht}%
  \BibitemOpen
  \bibfield  {author} {\bibinfo {author} {\bibfnamefont {B.}~\bibnamefont {Barman}}, \bibinfo {author} {\bibfnamefont {M.~R.}\ \bibnamefont {Haque}}, \ and\ \bibinfo {author} {\bibfnamefont {O.}~\bibnamefont {Zapata}},\ }\href {\doibase 10.1088/1475-7516/2024/09/020} {\bibfield  {journal} {\bibinfo  {journal} {JCAP}\ }\textbf {\bibinfo {volume} {09}},\ \bibinfo {pages} {020} (\bibinfo {year} {2024}{\natexlab{b}})},\ \Eprint {http://arxiv.org/abs/2405.15858} {arXiv:2405.15858 [astro-ph.CO]} \BibitemShut {NoStop}%
\bibitem [{\citenamefont {Zantedeschi}\ and\ \citenamefont {Visinelli}(2024)}]{Zantedeschi:2024ram}%
  \BibitemOpen
  \bibfield  {author} {\bibinfo {author} {\bibfnamefont {M.}~\bibnamefont {Zantedeschi}}\ and\ \bibinfo {author} {\bibfnamefont {L.}~\bibnamefont {Visinelli}},\ }\href@noop {} {\  (\bibinfo {year} {2024})},\ \Eprint {http://arxiv.org/abs/2410.07037} {arXiv:2410.07037 [astro-ph.HE]} \BibitemShut {NoStop}%
\bibitem [{\citenamefont {Dondarini}\ \emph {et~al.}(2025)\citenamefont {Dondarini}, \citenamefont {Marino}, \citenamefont {Panci},\ and\ \citenamefont {Zantedeschi}}]{Dondarini:2025ktz}%
  \BibitemOpen
  \bibfield  {author} {\bibinfo {author} {\bibfnamefont {A.}~\bibnamefont {Dondarini}}, \bibinfo {author} {\bibfnamefont {G.}~\bibnamefont {Marino}}, \bibinfo {author} {\bibfnamefont {P.}~\bibnamefont {Panci}}, \ and\ \bibinfo {author} {\bibfnamefont {M.}~\bibnamefont {Zantedeschi}},\ }\href@noop {} {\  (\bibinfo {year} {2025})},\ \Eprint {http://arxiv.org/abs/2506.13861} {arXiv:2506.13861 [hep-ph]} \BibitemShut {NoStop}%
\bibitem [{\citenamefont {Tan}\ and\ \citenamefont {Zhou}(2025)}]{Tan:2025vxp}%
  \BibitemOpen
  \bibfield  {author} {\bibinfo {author} {\bibfnamefont {X.-h.}\ \bibnamefont {Tan}}\ and\ \bibinfo {author} {\bibfnamefont {Y.-f.}\ \bibnamefont {Zhou}},\ }\href@noop {} {\  (\bibinfo {year} {2025})},\ \Eprint {http://arxiv.org/abs/2505.19857} {arXiv:2505.19857 [astro-ph.CO]} \BibitemShut {NoStop}%
\bibitem [{\citenamefont {Montefalcone}\ \emph {et~al.}(2025)\citenamefont {Montefalcone}, \citenamefont {Hooper}, \citenamefont {Freese}, \citenamefont {Kelso}, \citenamefont {Kuhnel},\ and\ \citenamefont {Sandick}}]{Montefalcone:2025akm}%
  \BibitemOpen
  \bibfield  {author} {\bibinfo {author} {\bibfnamefont {G.}~\bibnamefont {Montefalcone}}, \bibinfo {author} {\bibfnamefont {D.}~\bibnamefont {Hooper}}, \bibinfo {author} {\bibfnamefont {K.}~\bibnamefont {Freese}}, \bibinfo {author} {\bibfnamefont {C.}~\bibnamefont {Kelso}}, \bibinfo {author} {\bibfnamefont {F.}~\bibnamefont {Kuhnel}}, \ and\ \bibinfo {author} {\bibfnamefont {P.}~\bibnamefont {Sandick}},\ }\href@noop {} {\  (\bibinfo {year} {2025})},\ \Eprint {http://arxiv.org/abs/2503.21005} {arXiv:2503.21005 [astro-ph.CO]} \BibitemShut {NoStop}%
\bibitem [{\citenamefont {Dvali}\ \emph {et~al.}(2025)\citenamefont {Dvali}, \citenamefont {Zantedeschi},\ and\ \citenamefont {Zell}}]{Dvali:2025ktz}%
  \BibitemOpen
  \bibfield  {author} {\bibinfo {author} {\bibfnamefont {G.}~\bibnamefont {Dvali}}, \bibinfo {author} {\bibfnamefont {M.}~\bibnamefont {Zantedeschi}}, \ and\ \bibinfo {author} {\bibfnamefont {S.}~\bibnamefont {Zell}},\ }\href@noop {} {\  (\bibinfo {year} {2025})},\ \Eprint {http://arxiv.org/abs/2503.21740} {arXiv:2503.21740 [hep-ph]} \BibitemShut {NoStop}%
\bibitem [{\citenamefont {Dvali}\ and\ \citenamefont {Gomez}(2014)}]{Dvali:2013eja}%
  \BibitemOpen
  \bibfield  {author} {\bibinfo {author} {\bibfnamefont {G.}~\bibnamefont {Dvali}}\ and\ \bibinfo {author} {\bibfnamefont {C.}~\bibnamefont {Gomez}},\ }\href {\doibase 10.1088/1475-7516/2014/01/023} {\bibfield  {journal} {\bibinfo  {journal} {JCAP}\ }\textbf {\bibinfo {volume} {01}},\ \bibinfo {pages} {023} (\bibinfo {year} {2014})},\ \Eprint {http://arxiv.org/abs/1312.4795} {arXiv:1312.4795 [hep-th]} \BibitemShut {NoStop}%
\bibitem [{\citenamefont {Dvali}\ \emph {et~al.}(2017)\citenamefont {Dvali}, \citenamefont {Gomez},\ and\ \citenamefont {Zell}}]{Dvali:2017eba}%
  \BibitemOpen
  \bibfield  {author} {\bibinfo {author} {\bibfnamefont {G.}~\bibnamefont {Dvali}}, \bibinfo {author} {\bibfnamefont {C.}~\bibnamefont {Gomez}}, \ and\ \bibinfo {author} {\bibfnamefont {S.}~\bibnamefont {Zell}},\ }\href {\doibase 10.1088/1475-7516/2017/06/028} {\bibfield  {journal} {\bibinfo  {journal} {JCAP}\ }\textbf {\bibinfo {volume} {06}},\ \bibinfo {pages} {028} (\bibinfo {year} {2017})},\ \Eprint {http://arxiv.org/abs/1701.08776} {arXiv:1701.08776 [hep-th]} \BibitemShut {NoStop}%
\bibitem [{\citenamefont {Dvali}(2020)}]{Dvali:2020etd}%
  \BibitemOpen
  \bibfield  {author} {\bibinfo {author} {\bibfnamefont {G.}~\bibnamefont {Dvali}},\ }\href {\doibase 10.3390/sym13010003} {\bibfield  {journal} {\bibinfo  {journal} {Symmetry}\ }\textbf {\bibinfo {volume} {13}},\ \bibinfo {pages} {3} (\bibinfo {year} {2020})},\ \Eprint {http://arxiv.org/abs/2012.02133} {arXiv:2012.02133 [hep-th]} \BibitemShut {NoStop}%
\bibitem [{\citenamefont {Dvali}(2021{\natexlab{d}})}]{Dvali:2021bsy}%
  \BibitemOpen
  \bibfield  {author} {\bibinfo {author} {\bibfnamefont {G.}~\bibnamefont {Dvali}},\ }\href@noop {} {\  (\bibinfo {year} {2021}{\natexlab{d}})},\ \Eprint {http://arxiv.org/abs/2103.15668} {arXiv:2103.15668 [hep-th]} \BibitemShut {NoStop}%
\bibitem [{\citenamefont {Dvali}\ and\ \citenamefont {Gomez}(2013)}]{Dvali:2011aa}%
  \BibitemOpen
  \bibfield  {author} {\bibinfo {author} {\bibfnamefont {G.}~\bibnamefont {Dvali}}\ and\ \bibinfo {author} {\bibfnamefont {C.}~\bibnamefont {Gomez}},\ }\href {\doibase 10.1002/prop.201300001} {\bibfield  {journal} {\bibinfo  {journal} {Fortsch. Phys.}\ }\textbf {\bibinfo {volume} {61}},\ \bibinfo {pages} {742} (\bibinfo {year} {2013})},\ \Eprint {http://arxiv.org/abs/1112.3359} {arXiv:1112.3359 [hep-th]} \BibitemShut {NoStop}%
\bibitem [{\citenamefont {Alexandre}\ \emph {et~al.}()\citenamefont {Alexandre}, \citenamefont {Dvali}, \citenamefont {Turan},\ and\ \citenamefont {Zantedeschi}}]{Ana}%
  \BibitemOpen
  \bibfield  {author} {\bibinfo {author} {\bibfnamefont {A.}~\bibnamefont {Alexandre}}, \bibinfo {author} {\bibfnamefont {G.}~\bibnamefont {Dvali}}, \bibinfo {author} {\bibfnamefont {C.~J.}\ \bibnamefont {Turan}}, \ and\ \bibinfo {author} {\bibfnamefont {M.}~\bibnamefont {Zantedeschi}},\ }\href@noop {} {\enquote {\bibinfo {title} {Detectable mass range of primordial black holes with memory burden for theories with a large number of species},}\ }\bibinfo {note} {To appear}\BibitemShut {NoStop}%
\bibitem [{\citenamefont {Dvali}(2010{\natexlab{b}})}]{Dvali:2008fd}%
  \BibitemOpen
  \bibfield  {author} {\bibinfo {author} {\bibfnamefont {G.}~\bibnamefont {Dvali}},\ }\href {\doibase 10.1142/S0217751X10048895} {\bibfield  {journal} {\bibinfo  {journal} {Int. J. Mod. Phys. A}\ }\textbf {\bibinfo {volume} {25}},\ \bibinfo {pages} {602} (\bibinfo {year} {2010}{\natexlab{b}})},\ \Eprint {http://arxiv.org/abs/0806.3801} {arXiv:0806.3801 [hep-th]} \BibitemShut {NoStop}%
\bibitem [{\citenamefont {Dvali}\ and\ \citenamefont {Pujolas}(2009)}]{Dvali:2008rm}%
  \BibitemOpen
  \bibfield  {author} {\bibinfo {author} {\bibfnamefont {G.}~\bibnamefont {Dvali}}\ and\ \bibinfo {author} {\bibfnamefont {O.}~\bibnamefont {Pujolas}},\ }\href {\doibase 10.1103/PhysRevD.79.064032} {\bibfield  {journal} {\bibinfo  {journal} {Phys. Rev. D}\ }\textbf {\bibinfo {volume} {79}},\ \bibinfo {pages} {064032} (\bibinfo {year} {2009})},\ \Eprint {http://arxiv.org/abs/0812.3442} {arXiv:0812.3442 [hep-th]} \BibitemShut {NoStop}%
\bibitem [{Note2()}]{Note2}%
  \BibitemOpen
  \bibinfo {note} {The transition length scale for the many-copies model discussed below has been argued to be $R = ( M_{\protect \ensuremath {\protect \mathrm {P}}}^{2} / M_{\protect \ensuremath {\protect \mathrm {f}}}^{2} )^{1/n}\protect \hspace {0.5mm}M_{\protect \ensuremath {\protect \mathrm {f}}}^{-1}$ in Ref.~\cite {Dvali:2008fd}, exceeding the conservative upper bound in Eq.~\protect \eqref {eq:Micro-Black-Hole-Regime}.}\BibitemShut {Stop}%
\bibitem [{\citenamefont {Dvali}\ and\ \citenamefont {Redi}(2009)}]{Dvali:2009ne}%
  \BibitemOpen
  \bibfield  {author} {\bibinfo {author} {\bibfnamefont {G.}~\bibnamefont {Dvali}}\ and\ \bibinfo {author} {\bibfnamefont {M.}~\bibnamefont {Redi}},\ }\href {\doibase 10.1103/PhysRevD.80.055001} {\bibfield  {journal} {\bibinfo  {journal} {Phys. Rev. D}\ }\textbf {\bibinfo {volume} {80}},\ \bibinfo {pages} {055001} (\bibinfo {year} {2009})},\ \Eprint {http://arxiv.org/abs/0905.1709} {arXiv:0905.1709 [hep-ph]} \BibitemShut {NoStop}%
\bibitem [{\citenamefont {Argyres}\ \emph {et~al.}(1998)\citenamefont {Argyres}, \citenamefont {Dimopoulos},\ and\ \citenamefont {March-Russell}}]{Argyres:1998qn}%
  \BibitemOpen
  \bibfield  {author} {\bibinfo {author} {\bibfnamefont {P.~C.}\ \bibnamefont {Argyres}}, \bibinfo {author} {\bibfnamefont {S.}~\bibnamefont {Dimopoulos}}, \ and\ \bibinfo {author} {\bibfnamefont {J.}~\bibnamefont {March-Russell}},\ }\href {\doibase 10.1016/S0370-2693(98)01184-8} {\bibfield  {journal} {\bibinfo  {journal} {Phys. Lett. B}\ }\textbf {\bibinfo {volume} {441}},\ \bibinfo {pages} {96} (\bibinfo {year} {1998})},\ \Eprint {http://arxiv.org/abs/hep-th/9808138} {arXiv:hep-th/9808138} \BibitemShut {NoStop}%
\bibitem [{\citenamefont {Myers}\ and\ \citenamefont {Perry}(1986)}]{Myers:1986un}%
  \BibitemOpen
  \bibfield  {author} {\bibinfo {author} {\bibfnamefont {R.~C.}\ \bibnamefont {Myers}}\ and\ \bibinfo {author} {\bibfnamefont {M.~J.}\ \bibnamefont {Perry}},\ }\href {\doibase 10.1016/0003-4916(86)90186-7} {\bibfield  {journal} {\bibinfo  {journal} {Annals Phys.}\ }\textbf {\bibinfo {volume} {172}},\ \bibinfo {pages} {304} (\bibinfo {year} {1986})}\BibitemShut {NoStop}%
\bibitem [{\citenamefont {Myers}(1987)}]{Myers:1986rx}%
  \BibitemOpen
  \bibfield  {author} {\bibinfo {author} {\bibfnamefont {R.~C.}\ \bibnamefont {Myers}},\ }\href {\doibase 10.1103/PhysRevD.35.455} {\bibfield  {journal} {\bibinfo  {journal} {Phys. Rev. D}\ }\textbf {\bibinfo {volume} {35}},\ \bibinfo {pages} {455} (\bibinfo {year} {1987})}\BibitemShut {NoStop}%
\bibitem [{\citenamefont {Friedlander}\ \emph {et~al.}(2022)\citenamefont {Friedlander}, \citenamefont {Mack}, \citenamefont {Schon}, \citenamefont {Song},\ and\ \citenamefont {Vincent}}]{Friedlander:2022ttk}%
  \BibitemOpen
  \bibfield  {author} {\bibinfo {author} {\bibfnamefont {A.}~\bibnamefont {Friedlander}}, \bibinfo {author} {\bibfnamefont {K.~J.}\ \bibnamefont {Mack}}, \bibinfo {author} {\bibfnamefont {S.}~\bibnamefont {Schon}}, \bibinfo {author} {\bibfnamefont {N.}~\bibnamefont {Song}}, \ and\ \bibinfo {author} {\bibfnamefont {A.~C.}\ \bibnamefont {Vincent}},\ }\href {\doibase 10.1103/PhysRevD.105.103508} {\bibfield  {journal} {\bibinfo  {journal} {Phys. Rev. D}\ }\textbf {\bibinfo {volume} {105}},\ \bibinfo {pages} {103508} (\bibinfo {year} {2022})},\ \Eprint {http://arxiv.org/abs/2201.11761} {arXiv:2201.11761 [hep-ph]} \BibitemShut {NoStop}%
\bibitem [{\citenamefont {Cardoso}\ \emph {et~al.}(2006)\citenamefont {Cardoso}, \citenamefont {Cavaglia},\ and\ \citenamefont {Gualtieri}}]{Cardoso:2005mh}%
  \BibitemOpen
  \bibfield  {author} {\bibinfo {author} {\bibfnamefont {V.}~\bibnamefont {Cardoso}}, \bibinfo {author} {\bibfnamefont {M.}~\bibnamefont {Cavaglia}}, \ and\ \bibinfo {author} {\bibfnamefont {L.}~\bibnamefont {Gualtieri}},\ }\href {\doibase 10.1088/1126-6708/2006/02/021} {\bibfield  {journal} {\bibinfo  {journal} {JHEP}\ }\textbf {\bibinfo {volume} {02}},\ \bibinfo {pages} {021} (\bibinfo {year} {2006})},\ \Eprint {http://arxiv.org/abs/hep-th/0512116} {arXiv:hep-th/0512116} \BibitemShut {NoStop}%
\bibitem [{\citenamefont {Dvali}\ and\ \citenamefont {Gomez}(2010)}]{Dvali:2010vm}%
  \BibitemOpen
  \bibfield  {author} {\bibinfo {author} {\bibfnamefont {G.}~\bibnamefont {Dvali}}\ and\ \bibinfo {author} {\bibfnamefont {C.}~\bibnamefont {Gomez}},\ }\href@noop {} {\  (\bibinfo {year} {2010})},\ \Eprint {http://arxiv.org/abs/1004.3744} {arXiv:1004.3744 [hep-th]} \BibitemShut {NoStop}%
\bibitem [{Note3()}]{Note3}%
  \BibitemOpen
  \bibinfo {note} {The transition occurs essentially instantly for $p = 1$, but also for $p = 2$, one has $q(p = 2) \simeq 1 / \protect \sqrt {S}$, which is extremely early, given that $S \simeq 10^{30}\protect \,( M / 10^{10}\protect \,\protect \ensuremath {\protect \mathrm {g}})^{2}$.}\BibitemShut {Stop}%
\bibitem [{\citenamefont {Conley}\ and\ \citenamefont {Wizansky}(2007)}]{Conley:2006jg}%
  \BibitemOpen
  \bibfield  {author} {\bibinfo {author} {\bibfnamefont {J.~A.}\ \bibnamefont {Conley}}\ and\ \bibinfo {author} {\bibfnamefont {T.}~\bibnamefont {Wizansky}},\ }\href {\doibase 10.1103/PhysRevD.75.044006} {\bibfield  {journal} {\bibinfo  {journal} {Phys. Rev. D}\ }\textbf {\bibinfo {volume} {75}},\ \bibinfo {pages} {044006} (\bibinfo {year} {2007})},\ \Eprint {http://arxiv.org/abs/hep-ph/0611091} {arXiv:hep-ph/0611091} \BibitemShut {NoStop}%
\bibitem [{\citenamefont {Anchordoqui}\ \emph {et~al.}(2022)\citenamefont {Anchordoqui}, \citenamefont {Antoniadis},\ and\ \citenamefont {Lust}}]{Anchordoqui:2022txe}%
  \BibitemOpen
  \bibfield  {author} {\bibinfo {author} {\bibfnamefont {L.~A.}\ \bibnamefont {Anchordoqui}}, \bibinfo {author} {\bibfnamefont {I.}~\bibnamefont {Antoniadis}}, \ and\ \bibinfo {author} {\bibfnamefont {D.}~\bibnamefont {Lust}},\ }\href {\doibase 10.1103/PhysRevD.106.086001} {\bibfield  {journal} {\bibinfo  {journal} {Phys. Rev. D}\ }\textbf {\bibinfo {volume} {106}},\ \bibinfo {pages} {086001} (\bibinfo {year} {2022})},\ \Eprint {http://arxiv.org/abs/2206.07071} {arXiv:2206.07071 [hep-th]} \BibitemShut {NoStop}%
\bibitem [{\citenamefont {Anchordoqui}\ \emph {et~al.}(2023)\citenamefont {Anchordoqui}, \citenamefont {Antoniadis},\ and\ \citenamefont {Lust}}]{Anchordoqui:2022tgp}%
  \BibitemOpen
  \bibfield  {author} {\bibinfo {author} {\bibfnamefont {L.~A.}\ \bibnamefont {Anchordoqui}}, \bibinfo {author} {\bibfnamefont {I.}~\bibnamefont {Antoniadis}}, \ and\ \bibinfo {author} {\bibfnamefont {D.}~\bibnamefont {Lust}},\ }\href {\doibase 10.1016/j.physletb.2023.137844} {\bibfield  {journal} {\bibinfo  {journal} {Phys. Lett. B}\ }\textbf {\bibinfo {volume} {840}},\ \bibinfo {pages} {137844} (\bibinfo {year} {2023})},\ \Eprint {http://arxiv.org/abs/2210.02475} {arXiv:2210.02475 [hep-th]} \BibitemShut {NoStop}%
\bibitem [{\citenamefont {Anchordoqui}\ \emph {et~al.}(2024{\natexlab{a}})\citenamefont {Anchordoqui}, \citenamefont {Antoniadis},\ and\ \citenamefont {Lust}}]{Anchordoqui:2024akj}%
  \BibitemOpen
  \bibfield  {author} {\bibinfo {author} {\bibfnamefont {L.~A.}\ \bibnamefont {Anchordoqui}}, \bibinfo {author} {\bibfnamefont {I.}~\bibnamefont {Antoniadis}}, \ and\ \bibinfo {author} {\bibfnamefont {D.}~\bibnamefont {Lust}},\ }\href {\doibase 10.1103/PhysRevD.109.095008} {\bibfield  {journal} {\bibinfo  {journal} {Phys. Rev. D}\ }\textbf {\bibinfo {volume} {109}},\ \bibinfo {pages} {095008} (\bibinfo {year} {2024}{\natexlab{a}})},\ \Eprint {http://arxiv.org/abs/2401.09087} {arXiv:2401.09087 [hep-th]} \BibitemShut {NoStop}%
\bibitem [{\citenamefont {Anchordoqui}\ \emph {et~al.}(2024{\natexlab{b}})\citenamefont {Anchordoqui}, \citenamefont {Antoniadis},\ and\ \citenamefont {Lust}}]{Anchordoqui:2024dxu}%
  \BibitemOpen
  \bibfield  {author} {\bibinfo {author} {\bibfnamefont {L.~A.}\ \bibnamefont {Anchordoqui}}, \bibinfo {author} {\bibfnamefont {I.}~\bibnamefont {Antoniadis}}, \ and\ \bibinfo {author} {\bibfnamefont {D.}~\bibnamefont {Lust}},\ }\href {\doibase 10.1103/PhysRevD.110.015004} {\bibfield  {journal} {\bibinfo  {journal} {Phys. Rev. D}\ }\textbf {\bibinfo {volume} {110}},\ \bibinfo {pages} {015004} (\bibinfo {year} {2024}{\natexlab{b}})},\ \Eprint {http://arxiv.org/abs/2403.19604} {arXiv:2403.19604 [hep-th]} \BibitemShut {NoStop}%
\bibitem [{\citenamefont {Anchordoqui}\ \emph {et~al.}(2024{\natexlab{c}})\citenamefont {Anchordoqui}, \citenamefont {Antoniadis}, \citenamefont {Lust},\ and\ \citenamefont {Castillo}}]{Anchordoqui:2024jkn}%
  \BibitemOpen
  \bibfield  {author} {\bibinfo {author} {\bibfnamefont {L.~A.}\ \bibnamefont {Anchordoqui}}, \bibinfo {author} {\bibfnamefont {I.}~\bibnamefont {Antoniadis}}, \bibinfo {author} {\bibfnamefont {D.}~\bibnamefont {Lust}}, \ and\ \bibinfo {author} {\bibfnamefont {K.~P.~n.}\ \bibnamefont {Castillo}},\ }\href {\doibase 10.1016/j.dark.2024.101714} {\bibfield  {journal} {\bibinfo  {journal} {Phys. Dark Univ.}\ }\textbf {\bibinfo {volume} {46}},\ \bibinfo {pages} {101714} (\bibinfo {year} {2024}{\natexlab{c}})},\ \Eprint {http://arxiv.org/abs/2407.21031} {arXiv:2407.21031 [hep-th]} \BibitemShut {NoStop}%
\bibitem [{\citenamefont {Anchordoqui}\ \emph {et~al.}(2024{\natexlab{d}})\citenamefont {Anchordoqui}, \citenamefont {Antoniadis}, \citenamefont {Lust},\ and\ \citenamefont {Castillo}}]{Anchordoqui:2024tdj}%
  \BibitemOpen
  \bibfield  {author} {\bibinfo {author} {\bibfnamefont {L.~A.}\ \bibnamefont {Anchordoqui}}, \bibinfo {author} {\bibfnamefont {I.}~\bibnamefont {Antoniadis}}, \bibinfo {author} {\bibfnamefont {D.}~\bibnamefont {Lust}}, \ and\ \bibinfo {author} {\bibfnamefont {K.~P.~n.}\ \bibnamefont {Castillo}},\ }\href {\doibase 10.1016/j.dark.2024.101681} {\bibfield  {journal} {\bibinfo  {journal} {Phys. Dark Univ.}\ }\textbf {\bibinfo {volume} {46}},\ \bibinfo {pages} {101681} (\bibinfo {year} {2024}{\natexlab{d}})},\ \Eprint {http://arxiv.org/abs/2409.12904} {arXiv:2409.12904 [hep-ph]} \BibitemShut {NoStop}%
\bibitem [{\citenamefont {Anchordoqui}\ \emph {et~al.}(2025{\natexlab{a}})\citenamefont {Anchordoqui}, \citenamefont {Antoniadis},\ and\ \citenamefont {Lust}}]{Anchordoqui:2025nmb}%
  \BibitemOpen
  \bibfield  {author} {\bibinfo {author} {\bibfnamefont {L.}~\bibnamefont {Anchordoqui}}, \bibinfo {author} {\bibfnamefont {I.}~\bibnamefont {Antoniadis}}, \ and\ \bibinfo {author} {\bibfnamefont {D.}~\bibnamefont {Lust}},\ }\href@noop {} {\  (\bibinfo {year} {2025}{\natexlab{a}})},\ \Eprint {http://arxiv.org/abs/2501.11690} {arXiv:2501.11690 [hep-th]} \BibitemShut {NoStop}%
\bibitem [{\citenamefont {Anchordoqui}\ \emph {et~al.}(2025{\natexlab{b}})\citenamefont {Anchordoqui}, \citenamefont {Bedroya},\ and\ \citenamefont {L\"ust}}]{Anchordoqui:2025opy}%
  \BibitemOpen
  \bibfield  {author} {\bibinfo {author} {\bibfnamefont {L.~A.}\ \bibnamefont {Anchordoqui}}, \bibinfo {author} {\bibfnamefont {A.}~\bibnamefont {Bedroya}}, \ and\ \bibinfo {author} {\bibfnamefont {D.}~\bibnamefont {L\"ust}},\ }\href@noop {} {\  (\bibinfo {year} {2025}{\natexlab{b}})},\ \Eprint {http://arxiv.org/abs/2506.14874} {arXiv:2506.14874 [hep-ph]} \BibitemShut {NoStop}%
\bibitem [{\citenamefont {Anchordoqui}\ \emph {et~al.}(2025{\natexlab{c}})\citenamefont {Anchordoqui}, \citenamefont {Halzen},\ and\ \citenamefont {Lust}}]{Anchordoqui:2025xug}%
  \BibitemOpen
  \bibfield  {author} {\bibinfo {author} {\bibfnamefont {L.~A.}\ \bibnamefont {Anchordoqui}}, \bibinfo {author} {\bibfnamefont {F.}~\bibnamefont {Halzen}}, \ and\ \bibinfo {author} {\bibfnamefont {D.}~\bibnamefont {Lust}},\ }\href@noop {} {\  (\bibinfo {year} {2025}{\natexlab{c}})},\ \Eprint {http://arxiv.org/abs/2505.23414} {arXiv:2505.23414 [hep-ph]} \BibitemShut {NoStop}%
\bibitem [{\citenamefont {Ettengruber}\ \emph {et~al.}()\citenamefont {Ettengruber}, \citenamefont {K{\"u}hnel},\ and\ \citenamefont {Montefalcone}}]{Ettengruber-Constraints-2025}%
  \BibitemOpen
  \bibfield  {author} {\bibinfo {author} {\bibfnamefont {M.}~\bibnamefont {Ettengruber}}, \bibinfo {author} {\bibfnamefont {F.}~\bibnamefont {K{\"u}hnel}}, \ and\ \bibinfo {author} {\bibfnamefont {G.}~\bibnamefont {Montefalcone}},\ }\href@noop {} {}\bibinfo {note} {To appear}\BibitemShut {NoStop}%
\end{thebibliography}%
